\documentclass[12pt,apj]{emulateapj}
\usepackage{apjfonts}

\usepackage{color}
\newcommand{\HI}{\mbox{H\,\textsc{i}}}

\begin{document}

\shorttitle{Disk Warp Formation by Fly-by Encounters}
\shortauthors{Kim et al.}

\title{Formation of Warped Disks by Galactic Fly-by Encounters. I. Stellar Disks}
\author{Jeonghwan H. Kim\altaffilmark{1}, Sebastien Peirani\altaffilmark{2}, Sungsoo Kim\altaffilmark{3}, Hong Bae Ann\altaffilmark{4}, Sung-Ho An\altaffilmark{1}, and Suk-Jin Yoon\altaffilmark{1*}}

\affil{\altaffilmark{1}Department of Astronomy and Center for Galaxy Evolution Research, Yonsei University, Seoul 120-749, Korea\\ \altaffilmark{2}Institut d'Astrophysique de Paris (UMR 7095: CNRS and UPMC), 98 bis Bd Arago, 75014 Paris, France\\ \altaffilmark{3}Department of Astronomy and Space Science, Kyung Hee University, Yongin-shi, Kyungki 446-701, Korea\\ \altaffilmark{4}Division of Science Education, Pusan National University, Jangjeon-Dong Gumjeong-Gu, Busan 609-735, Korea}

\altaffiltext{*}{sjyoon@galaxy.yonsei.ac.kr}

\begin{abstract}
Warped disks are almost ubiquitous among spiral galaxies. 
Here we revisit and test the `fly-by scenario' of warp formation, 
in which impulsive encounters between galaxies are responsible for warped disks. 
Based on {\it N}-body simulations, we investigate the morphological and kinematical evolution 
of the stellar component of disks 
when galaxies undergo fly-by interactions with adjacent dark matter halos. 
We find that the so-called `S'-shaped warps can be excited by fly-bys 
and sustained for even up to a few billion years, 
and that this scenario provides a cohesive explanation for several key observations.
We show that disk warp properties are governed primarily by the following three parameters: 
(1) the impact parameter, i.e., the minimum distance between two halos,
(2) the mass ratio between two halos, and 
(3) the incident angle of the fly-by perturber. 
The warp angle is tied up with all three parameters, 
yet the warp lifetime is particularly sensitive to the incident angle of the perturber.
Interestingly, the modeled S-shaped warps are often non-symmetric depending on the incident angle.
We speculate that the puzzling U- and L-shaped warps are geometrically superimposed S-types 
produced by successive fly-bys with different incident angles, 
including multiple interactions with a satellite on a highly elongated orbit. 

\end {abstract}
\keywords{galaxies: evolution $-$ galaxies: halos $-$ galaxies: interactions $-$ galaxies: structure $-$ methods: numerical}

\section{Introduction}

The outer parts of most disk galaxies does not seem to be aligned with the inner planes of the disks, 
with warps apparent from an edge-on view \citep[e.g.,][]{1990MNRAS.246..458S,1998A&A...337....9R,2006NewA...11..293A}. 
There are two major types of warps: 
S-shaped warps, in which one side of the plane of the disk rises 
and the other side declines; 
and  U-shaped warps, in which
both sides of the plane rise. 
In addition to these two types, \citet{2003A&A...399..457S} introduced another type of warp, L-shaped warps, 
to describe a few galaxies with only a one-sided warp.

The first warped galaxy ever discovered was our Milky Way. 
\citet{1957AJ.....62...90B} and \citet{1957AJ.....62...93K} independently studied the shape of the Galaxy 
using 21 cm hydrogen-line observation 
and found that the maximum deviation of the plane exceeds 300 pc at a distance of 12 kpc from the Galactic center. 
Subsequent hydrogen-line observations of nearby edge-on galaxies confirmed 
that external galaxies also show significant warping at the outer gas regions \citep{1976A&A....53..159S}. 
Similar results were inferred from kinematic studies on less inclined galaxies 
\citep{1974ApJ...193..309R,1981AJ.....86.1791B,1981AJ.....86.1825B}. 
NGC 4013 is an example of a strong warp of an external galaxy 
\citep{1987Natur.328..401B,1995A&zA...295..605B,1996A&A...306..345B}.
Based on various \HI\ observations and parameterization of kinematics of massive spirals, 
it has been suggested that warps are rarer toward larger masses \citep{2007NewAR..51..120S}, 
and tend to be more asymmetric and have larger amplitudes in denser environments \citep{2002MNRAS.337..459G}. 
In addition, \citet{1990ApJ...352...15B} noted that warps are mainly noticeable at large radii 
where the optical image starts to diminish.

Stellar warps also exist \citep{1979A&AS...38...15V,1980ApJ...236L...1S,1987PASJ...39..849S}, 
although the amplitude of the warp angle seems smaller than that of \HI~\citep{2006NewA...11..293A}. 
There are some extreme cases of very strong stellar warps 
(even comparable with the strong radio warps) in interacting systems. 
One good example of a strong optical warp is in Mkn 305 along with Mkn 306 \citep{1990A&A...233..333K}. 
\citet{1990MNRAS.246..458S} showed that the warp frequency was higher than 80$\%$
of all northern-hemisphere NGC spiral galaxies.
Later, the same result was confirmed for the galaxies in the southern hemisphere \citep{2002A&A...391..519C}.
They also reported that no lenticular warped galaxy had been found.
\citet{1998A&A...337....9R} found evidence that more massive galaxies are less likely to warp, 
and most interacting galaxies show measurable warps, 
emphasizing the role of gravitational interaction. 
No correlation, however, has been found between the observed frequency of warped galaxies and spiral galaxy morphology.

To explain the formation and evolution of warped galaxies, 
several theoretical mechanisms have been proposed, 
including (1) intergalactic magnetic fields acting directly on the \HI\ gas in disks \citep{1990A&A...236....1B,1998A&A...332..809B}, 
(2) external torques originating from the gravitational forces of sinking satellites \citep{1997ApJ...480..503H,2013arXiv1307.5044S},
(3) discrete bending modes with the disk embedded in an axisymmetric halo \citep{1988MNRAS.234..873S},
(4) torques exerted by a misaligned halo \citep{1999ApJ...513L.107D,2000MNRAS.311..733I}, 
(5) direct accretion of intergalactic medium on disks \citep{2001ASPC..240..278R},
(6) reorientation of the outer parts of halos by cosmic infall \citep{1989MNRAS.237..785O,1999MNRAS.303L...7J,2006MNRAS.370....2S},
and (7) distortions in the dark matter halo by satellite galaxies \citep{2006ApJ...641L..33W}.

Many warped galaxies have nearby companions. 
This may support the idea that tidal interactions are involved in the creation of galaxy warps.
\citet{1972ApJ...178..623T} demonstrated the effects of close encounters on the galaxy evolution in great detail, 
although they focused mainly on the formation of galactic bridges and tails with various orbital parameters, 
not on galactic warps.
\citet{2000ApJ...534..598V} investigated the effects of distortion 
produced in dark matter halos during fly-by encounters, 
and found that such distortion of the halo might account for the formation of lopsided and warped disks. 
Similar interpretations were drawn from observational investigations \citep[e.g.,][]{1999MNRAS.304..330S}.

In a $\Lambda$CDM universe, galaxy mergers are considered key to understanding 
the formation and evolution of halos due to their dramatic influence on galaxy morphology and star formation rate. 
For this reason, various researchers \citep[e.g.,][]{1998ApJ...495..139M,1999MNRAS.304..465M,1999ApJ...526..607D,1999MNRAS.307..162S,2008MNRAS.388L..10B,2009A&A...496...51P,2010MNRAS.405.2327P} have investigated mergers. 
In contrast, yet another class of galaxy interaction, the fly-by encounter with no mergers involved, 
has been discounted, although it could be as frequent as, or even surpass the frequency of the merger 
\citep{2003ApJ...582..141G,2012ApJ...751...17S}.

Galaxies in cluster environments travel at a relatively high speed \citep{2003ApJ...582..141G},
and experience a number of fly-bys with high probability. 
\citet{2012ApJ...751...17S} paid attention to close halo fly-bys by analyzing high resolution {\it N}-body simulations. 
They found that halos with masses above $10^{12} M_{\odot} $ in low redshift ($z \leq 3$) experience more than 100 fly-by encounters per Gyr,
although the number of fly-bys among massive galaxy pairs is relatively small. 
They noted that about 70 \% of fly-bys are one-time events between halos. 
S.H. An et al. (2014, in preparation) also performed cosmological simulations 
to investigate the key characteristics of fly-by interactions as functions of redshift and mass ratios. 
Their results demonstrate that the number of equal mass fly-bys in clusters is comparable to the number of major mergers. 

For the following reasons, we hypothesize that galaxy fly-by interactions are responsible for galaxy warps in certain cases.
First, galactic fly-by interactions are frequent \citep{2012ApJ...751...17S}.
Furthermore, a few interesting relations reinforce the idea of a gravitational origin of warped disks:
(1) most interacting galaxies are warped and have greater warp amplitudes on average than isolated galaxies
\citep{1990A&A...233..333K,1998A&A...337....9R,2001A&A...373..402S},
(2) galaxies that have distinct tidal features show large warp asymmetries \citep{2006NewA...11..293A}, and
(3) more massive galaxies are somewhat less warped than less massive galaxies \citep{2007NewAR..51..120S}.

The main goal of this paper is to study whether a one-time, fly-by encounter with a perturber 
can generate a warp, with the structure being maintained for a long period of time.
To scrutinize the potential ramifications of galactic fly-by encounters, we utilize {\it N}-body simulations. 
In particular, we use a live set of halo {\it N}-body particles to produce a dark matter halo.
We prefer {\it live} halos to static fixed potential wells
so as not to underestimate the influence of interactions among halos
when halos overlap during the encounter.
We confine our discussion to the warp phenomenon of {\it stellar} disks, 
leaving the discussion on extended \HI\ warps to forthcoming papers in this series.

This paper is organized as follows.
In Section 2, we describe model galaxy construction and fly-by simulation set-ups for various parameters.
In Section 3, we present the methods we use to analyze simulated warps. 
Results for galaxies both in isolation and with encounters are shown in Section 4.
In Section 5, we summarize our findings and discuss the results.
We present conclusions in Section 6.

\section{Models}

\subsection{Construction of Model Galaxies}

\begin{deluxetable}{ccccc}
\tabletypesize{\scriptsize}
\tablecaption{Fraction, Mass, Number of Particles, and Particle Mass of Each Component for the Disk Galaxy Host\label{tbl-1}}
\tablewidth{0pt}
\tablehead{
\colhead{}&\colhead{$f^{\,a}$} &\colhead{$M_{tot}\,^{b}$} & \colhead{$N^{\,c}$} & \colhead{$PM^{\,d}$}\\[1pt]
&&$(M_{\odot})$&&$(M_{\odot})$
}
\startdata
DM & $0.94533$ & $9.003\times10^{11} $ & $1280935$ & $7.029\times10^{5}$\\
Disk & $0.0369$ & $3.514\times10^{10} $ & $100000$ & $3.514\times10^{5}$\\
Gas &$0.0041$ & $3.905\times10^{9} $ & $11111$ & $3.514\times10^{5}$\\
Bulge & $0.01367$ & $1.302\times10^{10} $ & $37046$ & $3.514\times10^{5}$ 
\enddata
\tablenotetext{a}{Mass fraction of each component.}
\tablenotetext{b}{Total mass of each component.}
\tablenotetext{c}{Number of particles.}
\tablenotetext{d}{Particle mass.}
\end{deluxetable}

\begin{deluxetable}{ccccc}
\tabletypesize{\scriptsize}
\tablecaption{Mass, Number of Particles and Particle Mass of Each Component for the Dark Matter Perturbers\label{tbl-2}}
\tablewidth{0pt}
\tablehead{
\colhead{Model}&\colhead{$M^{\,a}$} & \colhead{$N^{\,b}$} & \colhead{$PM^{\,c}$}\\[1pt]
&$(M_{\odot})$&&$(M_{\odot})$
}
\startdata
DM1 & $1.587\times10^{11} $ & $225835$ & $7.029\times10^{5}$\\
DM2 & $2.381\times10^{11} $ & $338753$ & $7.029\times10^{5}$\\
DM3 & $4.762\times10^{11} $ & $677506$ & $7.029\times10^{5}$ \\
DM4 & $9.524\times10^{11} $ & $1355013$ & $7.029\times10^{5}$\\
DM5 & $1.904\times10^{12} $ & $2710025$ & $7.029\times10^{5}$\\
DM6 & $3.809\times10^{12} $ & $5420050$ & $7.029\times10^{5}$
\enddata
\tablenotetext{a}{Mass of perturbers.}
\tablenotetext{b}{Number of particles.}
\tablenotetext{c}{Particle mass.}
\tablecomments{A perturber only consists of dark matter particles to avoid model complexity.}
\end{deluxetable}

Our aim is to investigate gravitational interactions between a disk galaxy (host) and a fly-by galaxy (perturber) 
using an idealized {\it N}-body + SPH simulation.  
Both objects are created following the prescription of \citet{2005MNRAS.361..776S}.
We use Gadget2 \citep{2005MNRAS.364.1105S} to perform fly-by simulations after construction of both the host and the perturber.

In particular, the host galaxy consists of a spherical dark matter halo with a Hernquist profile \citep{1990ApJ...356..359H}, 
a disk with an exponential surface density profile containing both stars and gas, and a bulge. 
The disk and bulge comprise 4.1\,\% and 1.4\,\% of the total mass respectively (Tables \ref{tbl-1}). 
The baryon fraction used here (5.46\,\%) is lower than the cosmic baryon fraction ($\sim$16.7\,\%) 
derived by \citet{2009ApJS..180..330K},
but consistent with that of \citet{2005MNRAS.361..776S}.
Observational analysis suggests that most galaxies are severely baryon-depleted 
with respect to the cosmological fraction \citep[see for instance][]{2003ApJS..149..289B,2005ApJ...635...73H,2010ApJ...708L..14M}. 
The gas fraction of the disk is 10\,\% in the host galaxy, consistent with \citet{2006A&A...447..113L}.
In our fiducial model the host has a mass of 9.52$\times10^{11} M_{\odot}$, 
which corresponds to a virial velocity ($v_{200}$) of 160 km s$^{-1}$ comparable with the Milky Way.
The values of the concentration index of the halo ({\it c}) and the spin parameter of the halo ($\lambda$)  are 9 and 0.033, respectively.
The disk scale length ($R_d$), which is effectively determined by the spin parameter value, is 2.34 kpc for the host galaxy in our case.

For the perturber, we only consider a dark matter halo (${\it c}=9, \,\lambda=0.033$) with the Hernquist profile 
because even for the closest passage, only halos are interpenetrated 
without direct interactions between baryonic components, i.e., bulges and disks. 
Six model perturbers with different masses, ranging from $1.587\times10^{11}\, M_{\odot}$ (DM1) to $3.809\times10^{12}\, M_{\odot}$ (DM6), are used.
Detailed information about the perturber models (their total masses, the number of particles, and the particle masses of each component) is listed in Table \ref{tbl-2}. 
Note that the particle mass of each component must match between the host and the perturber.

\subsection{Initial Parameters for Fly-by Encounters}
\begin{figure}[t]

\centerline{\includegraphics[width=\linewidth]{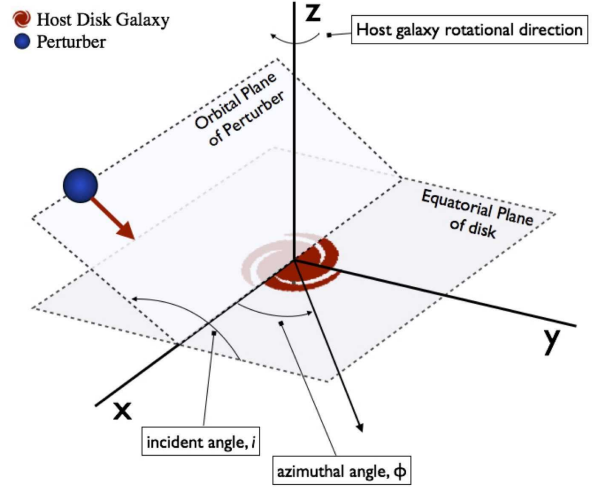}}

\caption[]{Schematic view of fly-by simulations. The host (red) is placed at the origin rotating in the {\it x-y} plane. The perturber (blue) is located 600 kpc away from the origin with a relative velocity of 600 km s$^{-1}$ and an incident angle $i$. The incident angle is the angle between the equatorial plane of the host and the orbital plane of the perturber.}
\label{sview}
\end{figure}

Once we build host and perturbing galaxies, their initial positions, initial velocities, and orientation angles are defined. 
A schematic view of the fly-by simulations is provided in Figure \ref{sview}. 
For the purpose of this study, one disk galaxy is placed at the origin with no initial velocity 
while one dark halo perturber is located 600 kpc away from the host 
with a relative velocity of 600 km s$^{-1}$.
The host disk galaxy rotates in the {\it x-y} plane.
The azimuthal angle is measured from the x-axis to quantify the evolution of tips of warps.
\citet{2003ApJ...582..141G}, based on a Virgo-type cluster simulation, 
found that the relative velocity of galaxies at encounters peaks at $\sim$ 350 km s$^{-1}$ 
and has a mean value of around 800 km s$^{-1}$ showing a skewed distribution. 
In addition, \citet{1998MNRAS.299..728T} reported that galaxies have a peak relative velocity of $\sim$ 500 km s$^{-1}$. 
Our initial relative velocity of 600 km s$^{-1}$ lies in the range of the values previously proposed, 
and is therefore an appropriate representative value.
We note that the total tidal strength exerted on the galaxy 
is proportional to the integration time (i.e., the duration) of the encounter. 
Accordingly, interactions with smaller relative velocities
have a higher chance of triggering warp formation.

Each simulation run has a different impact parameter, mass ratio, or incident angle of the perturber.
The impact parameter ($R_{ip}$) is the minimum distance between the host and the perturber at the moment of the minimum distance ($t_{ip}$).
The mass ratio is defined as
\begin{equation}
Mass{\,\,}Ratio=\frac{M_{perturber}}{M_{host}},
\end{equation}
such that a higher mass ratio indicates a more massive perturber 
with respect to the host galaxy.
The incident angle ($i$) denotes the angle of the perturber's orbit to the rotational plane of the host.
In other words, we set the incident angle to $0^{\circ}$ if the perturber moves along the plane in the same direction as disk rotation (prograde passage).
By convention, $i$ does not exceed $180^{\circ}$.
The configurations of all runs are listed in Table \ref{tbl-3}.

\begin{deluxetable}{cccccc}
\tabletypesize{\tiny}
\tablecaption{Configuration of the Fly-by Encounter Simulations\label{tbl-3}}
\tablewidth{0pt}
\tablehead{
\colhead{Run$^a$}&\colhead{Perturber$^{\,b}$}& \colhead{$R_{ip}\,^{c}$}& \colhead{$M_{p}/M_{h}\,^d$}  & \colhead{$i^{\,e}$} & \colhead{$t_{ip}\,^f$}\\
& &(kpc) &  &($^{\circ}$)& (Gyr)
}
\startdata
IP1& DM4 & $33.14$ & $1.0$ &  $90$ & $0.96$\\
IP2& DM4 & $42.88$ & $1.0$ & $90$ & $0.96$\\
IP3& DM4 & $51.91$ & $1.0$ & $90$ & $0.96$\\
IP4& DM4 & $60.63$ & $1.0$ & $90$ & $0.96$\\
IP5& DM4 & $69.69$ & $1.0$ & $90$ & $0.96$\\
IP6& DM4 & $79.05$ & $1.0$ & $90$ & $0.96$\\
IP7& DM4 & $88.48$ & $1.0$ & $90$ & $0.96$\\
IP8& DM4 & $109.56$ & $1.0$ & $90$ & $0.96$\\
IP9& DM4 & $138.87$ & $1.0$ & $90$ & $0.96$\\
IP10& DM4 & $187.54$ & $1.0$ & $90$ & $0.94$\\
\hline
M1& DM1 & $45.61$ & $0.17$ & $90$ & $0.97$\\
M2& DM2 & $45.47$ & $0.25$ & $90$ & $0.97$\\
M3& DM3 & $44.34$ & $0.5$ & $90$ & $0.97$\\
M4& DM4 & $42.88$ & $1.0$ & $90$ & $0.96$\\
M5& DM5 & $43.13$ & $2.0$ & $90$ & $0.94$\\
M6& DM6 & $41.83$ & $4.0$ & $90$ & $0.92$\\
\hline
A1& DM4 & $42.83$ & $1.0$ &  $0$ & $0.96$\\
A2& DM4 & $42.06$ & $1.0$ & $30$ & $0.96$\\
A3& DM4 & $41.94$ & $1.0$ & $45$ & $0.96$\\
A4& DM4 & $42.05$ & $1.0$ & $60$ & $0.96$\\
A5& DM4 & $42.88 $ & $1.0$ & $90$ & $0.96$\\
A6& DM4 & $42.05$ & $1.0$ & $120$ & $0.96$\\
A7& DM4 & $42.21$ & $1.0$ & $135$ & $0.96$\\
A8& DM4 & $42.05$ & $1.0$ & $150$ & $0.96$\\
A9& DM4 & $42.74$ & $1.0$ & $180$ & $0.96$
\enddata
\tablenotetext{a}{Name of runs.}
\tablenotetext{b}{Model perturbers.}
\tablenotetext{c}{Impact parameter : Minimum distance between the host and perturber.}
\tablenotetext{d}{Mass ratio (perturber/host).}
\tablenotetext{e}{Angle between the equatorial plane of the host galaxy and the orbital plane of perturber.}
\tablenotetext{f}{Time when $R$ is closest.}
\end{deluxetable}

\section{Analysis}

\subsection{Angle Measurements}

\begin{figure}[t]

\centerline{\includegraphics[width=\linewidth]{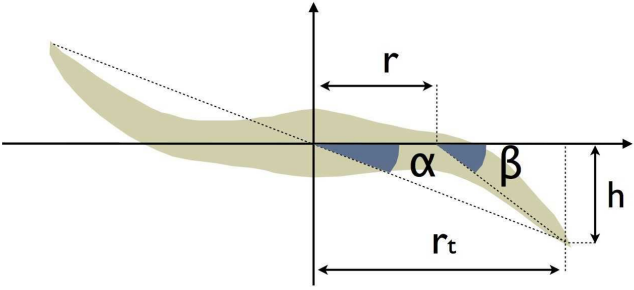}}

\caption[Definition of warp angles]{Definition of the warp angles $\alpha$ and the angle $\beta$ 
for a schematic edge-on warped disk (thick gray object). 
$r$ and $r_t$ are the warp's starting radius 
and the projected distance along the major axis to the last measured point, respectively.}
\label{warpangle}
\end{figure}

Figure \ref{warpangle} shows the definitions of the warp angles, $\alpha$ and $\beta$. 
The angle $\alpha$ is the angle between the major axis of a galaxy 
and the outermost point of the disk, and is given by 
\begin{equation}
\tan{\alpha}=\frac{h}{r_t}\,\,\,,
\end{equation}
where {\it h} is the distance of the tip of the outermost point from the disk major axis
and $r_t$ is the distance of the tip of the outermost point from the galaxy center along the {\it x}-axis (i.e., disks major axis).
The angle $\beta$ is the angle between the galaxy's major axis 
and the line drawn from the point where the warp becomes evident 
to the outermost point of the disk, and is given by 
\begin{equation}
\tan{\beta}=\frac{h}{r_t - r}\,\,\,,
\end{equation}
where {\it r} is the warp starting radius.
In this scheme, the warp angle $\alpha$ is smaller than the angle $\beta$. 

As shown in Figure \ref{warpangle}, evaluating $\beta$ accurately is closely related to measuring $r$. 
It is clear that errors associated with the measurement of $\beta$ are larger than those of $\alpha$ 
because of ambiguity in defining the warp starting point {\it r} of the disks. 
For this reason, we use the warp angle $\alpha$ rather than $\beta$
in the analysis of our simulated galaxies throughout this study.

\subsection{Ring Model}

To quantify the warp structure we use the tilted-ring model, 
which was first introduced by \citet{1974ApJ...193..309R} to explain M83's warp.
\citet{1981AJ.....86.1791B} and \citet{1997MNRAS.292..349S} have shown 
that the orbits of material within spiral galaxies have low ellipticity. 
Thus, treating disk material as being circular has validity. 
To measure warp angles, each simulated galaxy is divided into 10 successive rings, or annuli, 
of equal width from the center of galaxy out to 20 kpc in radius. 
For example, the first ring represents the inner region of the disk from its center to 2 kpc in radius 
while the final tenth annuli is for the outermost parts of the disk between 18 and 20 kpc in radius. 
By dividing a disk into equal width, we can calculate
the warp angle for each annuli, the warp starting points, and their sensitivity to fly-by encounters.

Warp is in general stronger at the outer part of disks 
and thus the outer rings in the simulation better represent the structure. 
However, the outermost region contains only a small fraction of the mass of a disk or the total number of particles. 
Hence, we only consider regions up to 20 kpc in radius 
to avoid the statistical uncertainty caused by an insufficient number of particles.

\subsection{Definition of Warp Lifetime, $t_{L}$}

Estimating the lifetime of a warp is crucial 
because lifetime is directly correlated with the probability of a warp being observed. 
To estimate the lifetime of warps from simulations, 
we adopt the following simple approach. 
Warp angles are recorded for 
the isolated system, i.e., an unperturbed galaxy acting as a control sample. 
Throughout the paper, we postulate that a galaxy is warped 
only if the warp angle exceeds two times the standard deviation (2$\sigma$)
of the averaged warp angle of the isolated, unperturbed galaxy disk. 
To model the pattern of change in the warp angle with time elapsed after an encounter, 
we apply polynomial fitting to our resultant data. 
A fifth order polynomial fit provides the best model for the warp evolution pattern. 
As our simulations are only performed for 5 Gyr, 
we use additional linear fits to the data for warps that last longer than 5 Gyr
to obtain the extrapolated lifetime.

\begin{figure*}[]
\centerline{\includegraphics[width=\linewidth]{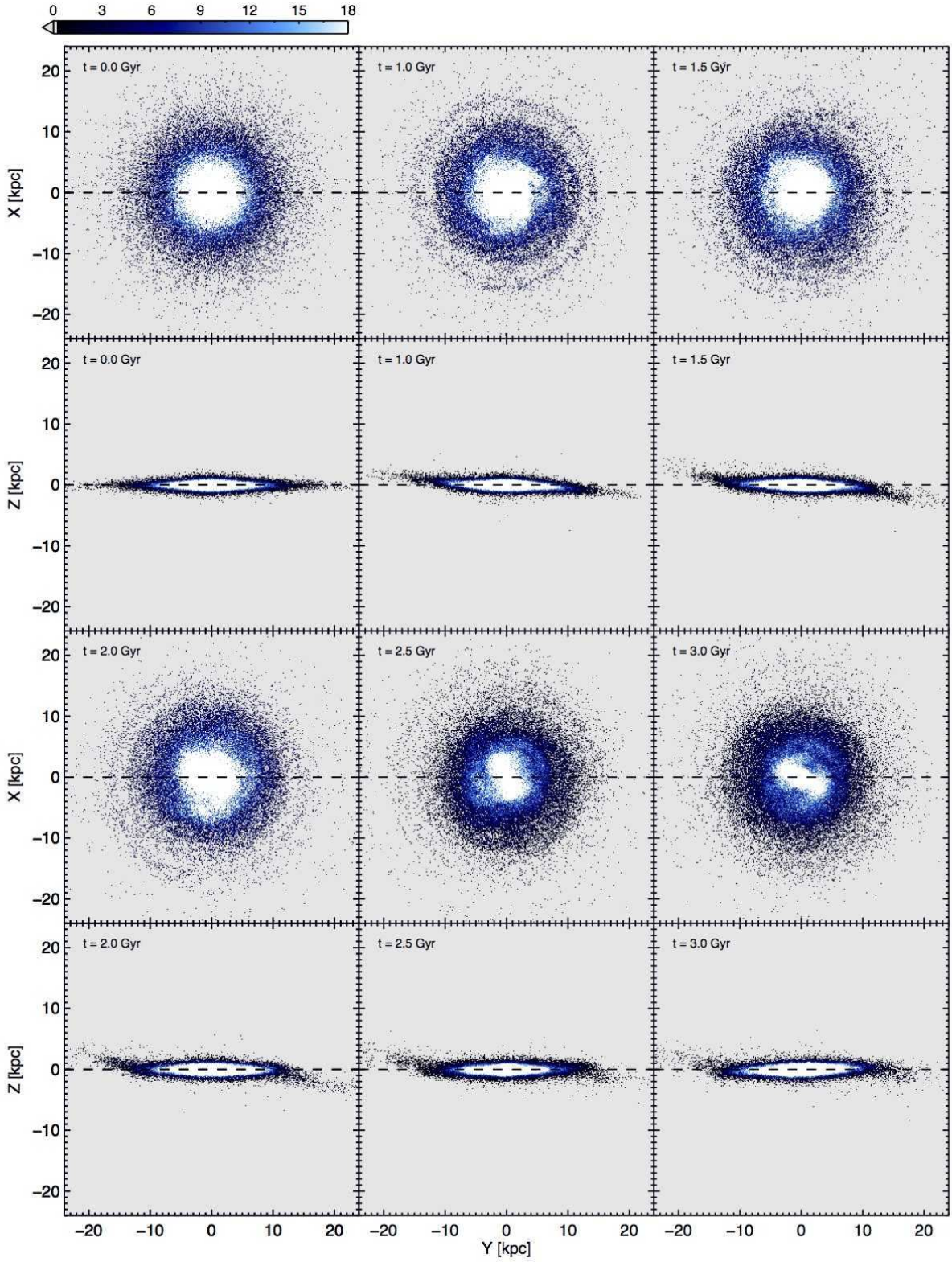}}
\caption[]{Evolution of a warped galaxy due to fly-by interaction with a dark halo (run A8). 
The projected stellar particle number density is shown at different epochs. 
The top-left color bar displays the level of the number density where all zero values are plotted in gray. 
The galaxy rotates counterclockwise.}
\label{ex}
\end{figure*}

\begin{figure*}[htbp]
\centerline{\includegraphics[width=\linewidth]{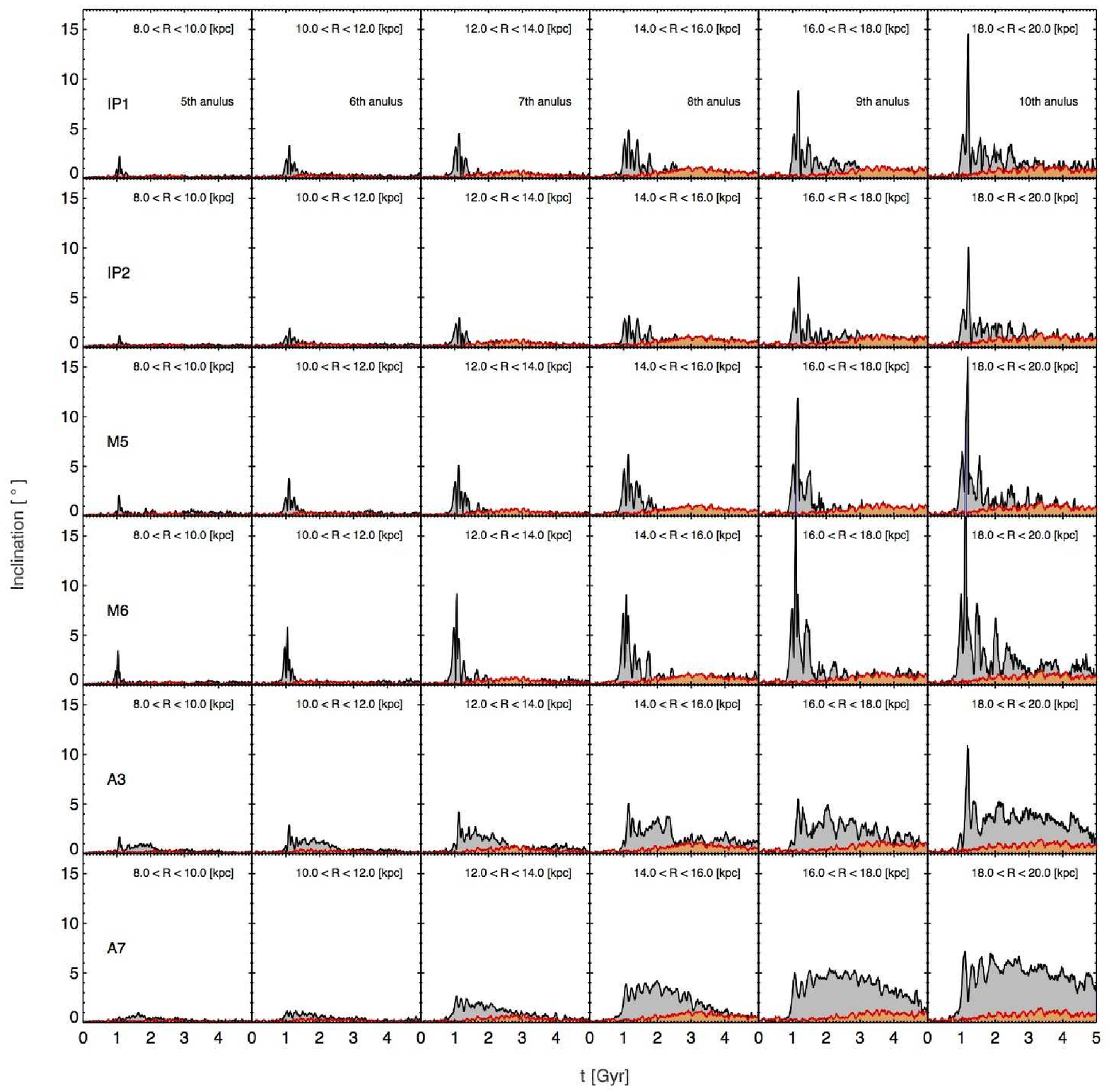}}
\caption[]{Evolution of the warp inclination angle $\alpha$ for each annulus. 
These selected simulation runs are examples that show a distinct sign of warps. 
In each panel, the red line displays the evolution of the warp amplitude for an isolated, unperturbed disk.
Each row, a total of six panels, forms a set of simulation runs with each ring or annuli presented.}
\label{represent_fit}
\end{figure*}

\section{Results}
\subsection{Isolated Disk Galaxy as a Control Sample}

Simulation of an isolated, unperturbed system shows 
no or little warp at the inner rings ({\it R} \,$\lesssim$\,10 kpc),
but a very weak warp ($\lesssim 1^{\circ}$) at the outer rings ({\it R} \,$\gtrsim$\,10 kpc). 
For example, the outer regions of the disk, represented by Rings 8$-$10,
have noticeable variations in inclination angle compared to the stable inner regions represented by Ring 5$-$7.
In our working hypothesis, warps are not expected for isolated galaxies
and we thus attribute the weak warps to random fluctuation of modeled rings 
mainly due to the lack of particles residing in each annulus. 
Ring 10, for instance, only contains 0.2 $\%$ of the total mass of the galaxy disk.

Hereafter, the change in the inclination angle of an isolated galaxy will be used 
as the background level of the real warp angle for all simulations performed. 
We identify a disk as being warped only if the warp angle $\alpha$ exceeds two times the standard deviation (2$\sigma$) 
of average inclination angles of the isolated system in each bin.

\begin{figure*}[htbp]
\centerline{\includegraphics[width=0.95\linewidth]{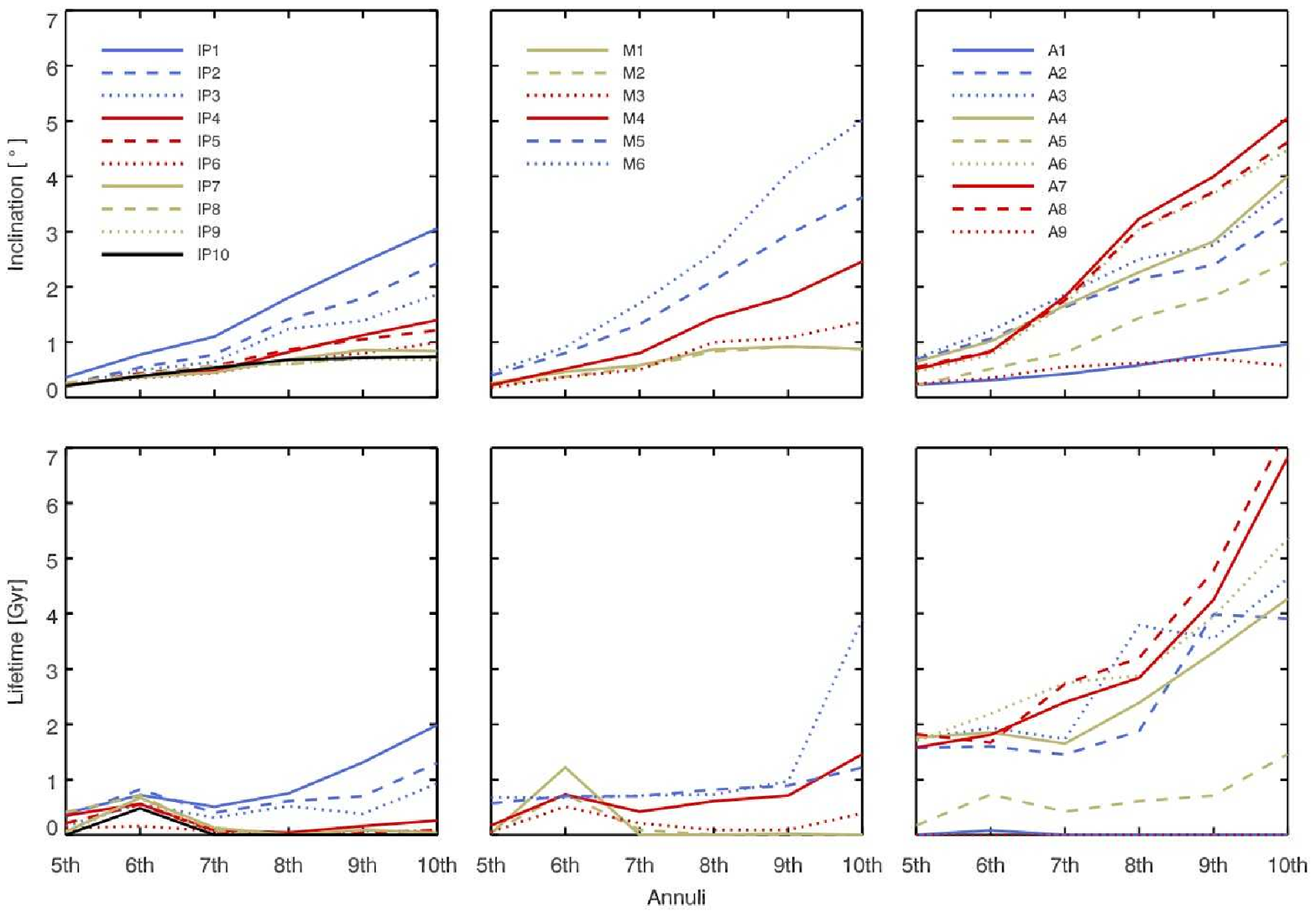}}
\caption[]{Variation of the averaged inclination angle (upper panels) and lifetime (lower) for each annulus. 
The inclination angles are averaged over a 1 Gyr period (between $t=0$ Gyr and $t=2$ Gyr).
Ring numbers are depicted in the {\it x}-axis.
The left, middle, and right columns are for different impact parameters (`IP' runs), 
mass ratios (`M' runs), and incident angles (`A' runs), respectively.}
\label{lifeinc}
\end{figure*}

\subsection{Disk Galaxies after Fly-by Encounters}

An example of a warped galaxy after a fly-by interaction with a dark halo is illustrated in Figure \ref{ex}. 
As the perturber gets closer to the host, the near side of the galaxy starts to bend ($t<1$),
and after the perturber passes by ($t\ge1$), a warp develops on the further side, resulting in an integral-shaped warp.
The outermost regions maintain the warp for a few billion years (depending on models), 
but warps in the inner regions disappear quickly.
The size of the warp angles of our model galaxies ranges from $\alpha$ = 2$^{\circ}$ $-$ 5$^{\circ}$,
consistent with observations \citep{1995A&AT....8...31R,1998A&A...337....9R,2006NewA...11..293A}.
The present study considers three parameters---the impact parameter $R_{ip}$, mass ratio $M_p/M_h$, 
and incident angle $i$ (Table \ref{tbl-3}).  
Figures \ref{represent_fit} and \ref{lifeinc} demonstrate the effects of these parameters. 
We show the evolution of the warp amplitude for an isolated, unperturbed disk in Figure \ref{represent_fit} (red region)
to compare with the evolution of a disk galaxy with and without a fly-by encounter with a dark matter perturber.

\subsubsection{The Effect of the Impact Parameter}

We perform a total of 10 simulation runs (Table \ref{tbl-3}, `IP' runs) 
to explore the effect of the impact parameter, $R_{ip}$, on the creation of galaxy warp. 
The $R_{ip}$ values range from $\sim$\,30 kpc (IP1) to $\sim$\,190 kpc (IP10).
We do not perform simulations with $R_{ip}<$ 30 kpc 
to avoid destruction of the internal structures of disks.

The top row of Figure \ref{represent_fit} shows the result of the IP1 run, 
for which two galaxies have the minimum distance at $t_{ip}$ 
and thus the host experiences the strongest tidal force among all IP runs. 
The perturber passes by the disk galaxy at $ t \sim 1$ Gyr
and the magnitude of the warp reaches its peak about 0.2 Gyr after the encounter. 
The minimum distance $R_{ip}$ between the disk galaxy and the perturber
is approximately 33 kpc at $t_{ip} = 0.96$ Gyr.
The system has a mass ratio ($M_{p}/M_{h}$) of 1 with an incident angle ($i$) of $90^{\circ}$. 
The IP1 run has the largest inclination angle
and the longest warp lifetime among all IP runs.
Warping of the outer parts is more pronounced than that of the inner region. 
For instance, the inner Ring 5 has a maximum warp angle of about $2.2^{\circ}$. 
The amplitudes of Rings 6, 7, 8, and 9 have maximum angles of 
approximately $3.29^{\circ}$, $4.56^{\circ}$, $4.82^{\circ}$, and $8.85^{\circ}$, respectively. 
The outermost region of the disk, represented by Ring 10, 
has the largest warp amplitude ($>$\,$14^{\circ}$).

The IP2 run shown in the second row of Figure \ref{represent_fit} has an $R_{ip}$ value of $\simeq$ 42 kpc. 
Compared to the previous run (IP1), the amplitudes are lower, as expected. 
The inner region, Ring 5, has a small warp, 
while the outermost region, Ring 10, has a warp angle slightly above $10^{\circ}$. 
The warp lifetime in this case is shorter than that of the IP1 run.

The leftmost column of Figure \ref{lifeinc}  shows the averaged inclination over 1 Gyr 
and the lifetime of each IP run as functions of galactocentric distance (i.e., annulus number).
Runs IP3 to IP10 correspond to $R_{ip} \sim$ 51, 60, 69, 79, 88, 109, 138, and 187 kpc, respectively. 
As $R_{ip}$ increases, the warp amplitude $\alpha$ and the estimated lifetime $t_{L}$ decreases. 
These results suggest that a one-time, fly-by encounter with $M_{p}/M_{h} \leq$ 1 
can generate warp and that this structure can be sustained for a few billion years 
if $R_{ip}$  is close enough.

\begin{figure*}[htbp]
\includegraphics[width=\linewidth]{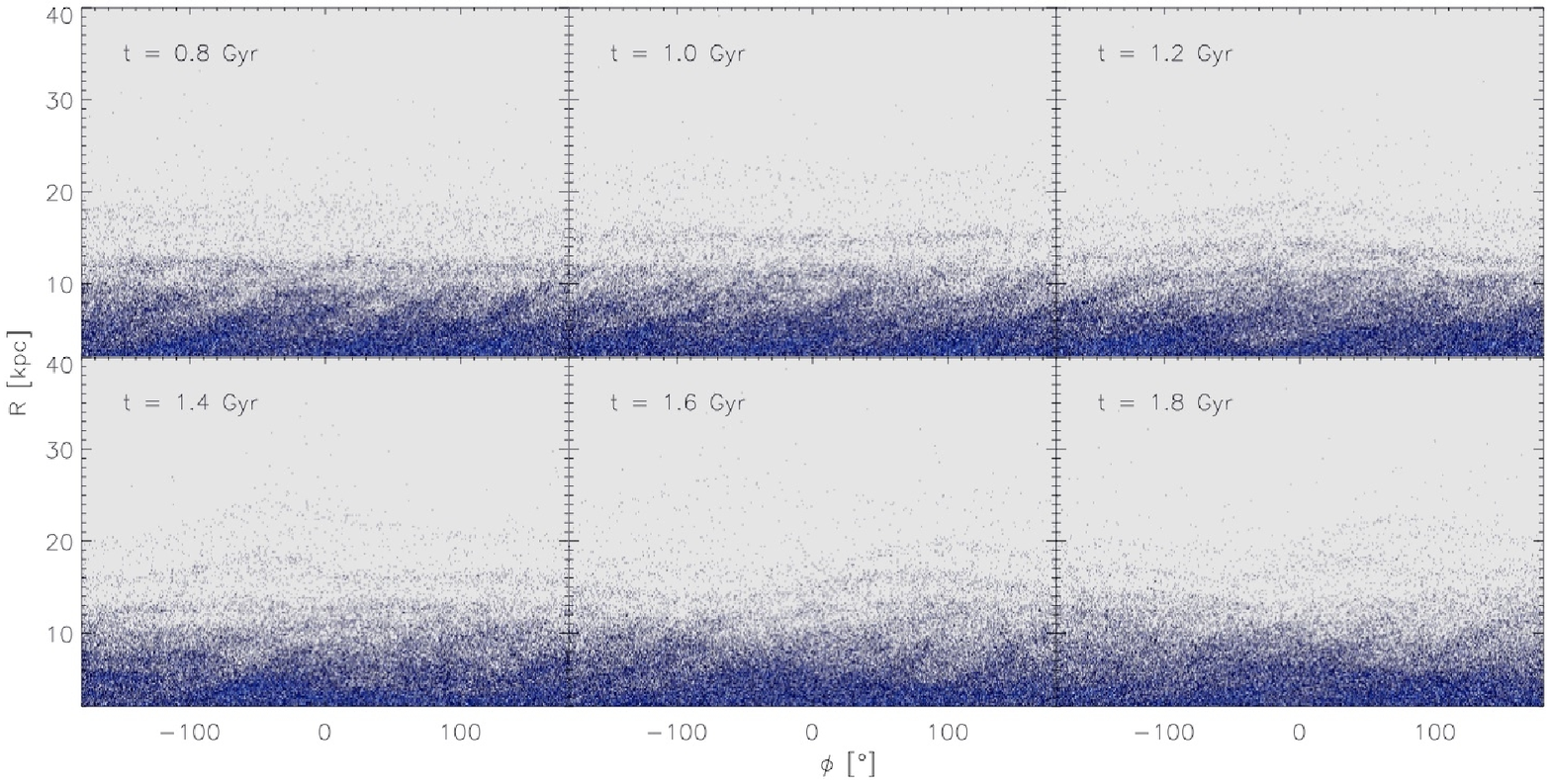}
\includegraphics[width=\linewidth]{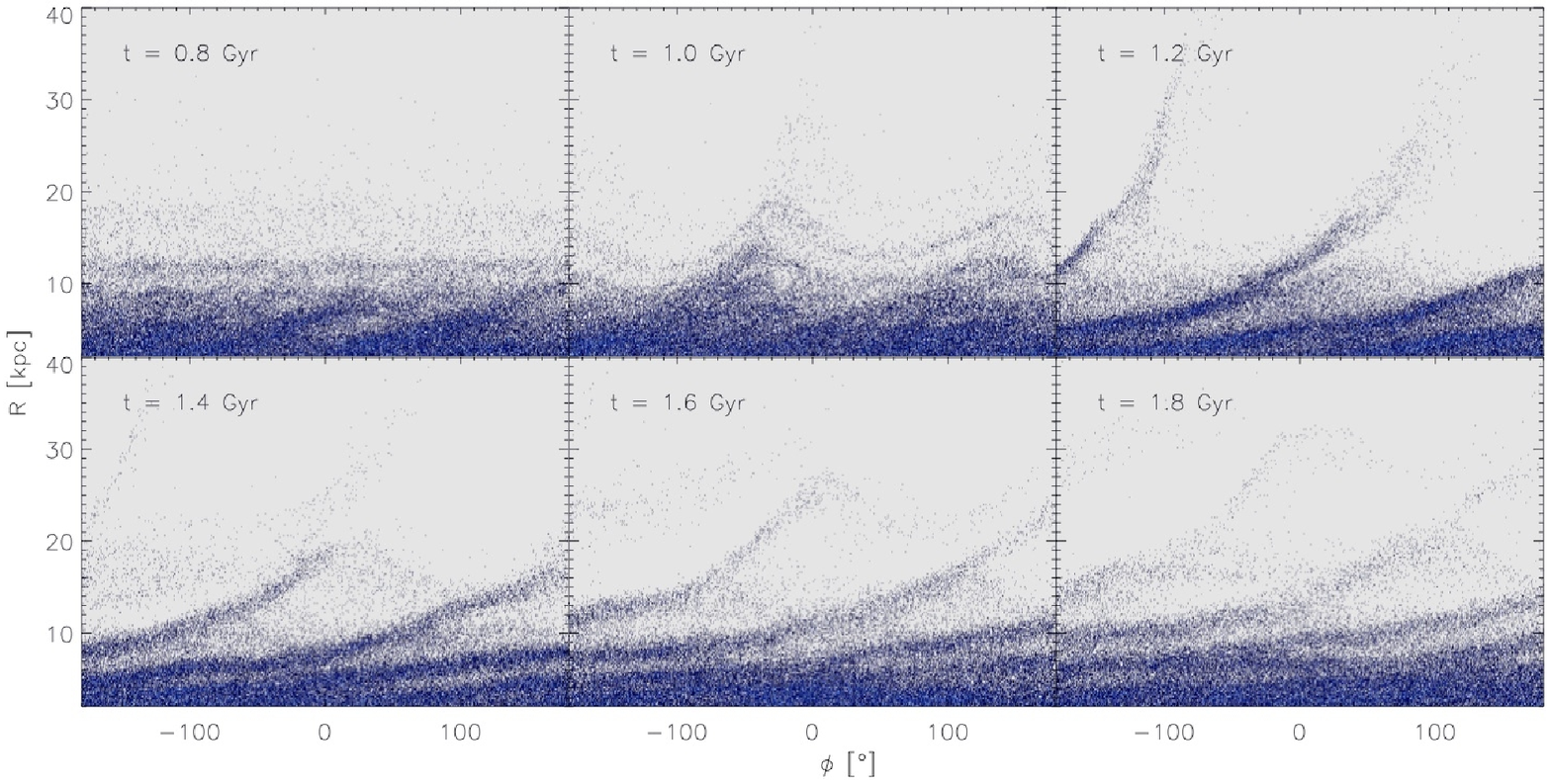}
\caption[]{Distribution of the disk particles of the A9 (top) and A1 (bottom) models. In each plot, the {\it x}- and {\it y}-axes represent the position angle $\phi$ (a position angle of 0 matches with {\it x}-axis in our schematic diagram in Figure \ref{sview}) and the distance from the galaxy center, respectively. The latter is the case in which the perturber moves in the same direction (prograde), exerting more gravitational force due to the increase in the duration of interaction. As a result, a distinct sign of tidal arms is shown in the bottom panels from {\it t} = 1.0 $-$ 2.0 Gyr.}
\label{surface_density}
\end{figure*}

\begin{figure*}[htbp]
\centerline{\includegraphics[width=\linewidth]{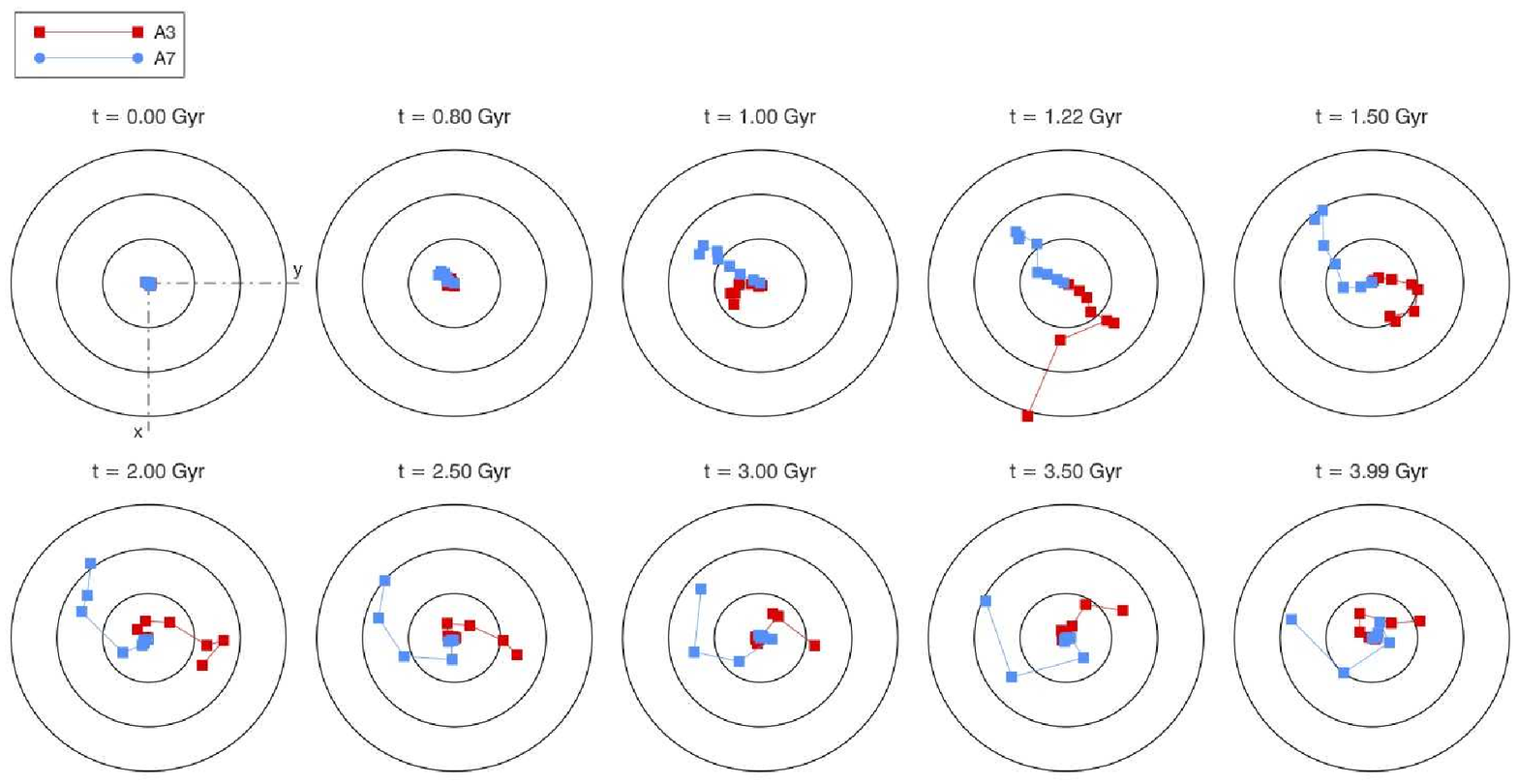}}
\caption[]{Briggs diagrams for the A3 (red) and A7 (blue) simulation runs at several different epochs. Each point represents a radial bin. There are a total of 10 bins spaced equally between 0 and 20 kpc. The polar coordinate is the line of nodes (LONs). The concentric circles are plotted in intervals of $3^{\circ}$. Direction of disk rotation is clockwise. The dashed-dotted lines in the first diagram at {\it x} = 0.0 Gyr match the direction of the {\it x}-axis and {\it y}-axis in Figure \ref{sview}. By convention, {\it x} and {\it y} correspond to $\phi = 0^{\circ}$ and $\phi = 90^{\circ}$.}
\label{pa}
\end{figure*}

\subsubsection{The Effect of the Mass Ratio}

We perform a total of six simulations (Table \ref{tbl-3}, `M' runs) 
to examine the effect of the mass ratio between the disk galaxy and the perturber. 
The $M_{p}/M_{h}$ values range from $\sim$\,1/6 (M1) to $\sim$\,4 (M6).

The third and fourth rows of Figure \ref{represent_fit} show the results of the M5 and M6 runs, 
for which the two galaxies have the largest $M_{p}/M_{h}$ ratios of 2 and 4, respectively,
and thus the hosts experience the strongest tidal force among all M runs. 
The systems have $R_{ip} \simeq 42$ kpc. 
For both runs, Rings 9 and 10 become highly warped, reaching a maximum angle of $>$ $15^{\circ}$
with warp lifetimes of $\geq 3 \sim 4$ Gyr.

The middle column of Figure \ref{lifeinc} shows the averaged inclination and lifetime
of each M run in terms of the annulus.
When the mass of the perturber is one-sixth  that of the disk galaxy (M1 run), 
no significant result is found, implying that it is unlikely to generate warps. 
Similarly, the interaction in run M2 with $M_{p}/M_{h} =$ 0.25 and $R_{ip}\simeq$ 45 kpc 
simply cannot generate a tidal force strong enough to excite warps. 
Therefore, a one-time fly-by encounter with $M_{p}/M_{h} \leq$ 0.25 and $R_{ip} \simeq $ 45 kpc 
cannot cause warps 
unless the galaxy pair has a very small relative velocity and thus a sufficient interaction duration.

When $M_{p}/M_{h}$ reaches 0.5 (the M3 run), however, the warps become apparent. 
In this case, the outer regions of galaxies are clearly warped ($\alpha \sim$ $2.2^{\circ}$ and \,$2.8^{\circ}$). 
Interactions with more massive perturbers ($M_{p}/M_{h} \geq $ 1) generate highly warped galaxies. 
In these simulations, the maximum warp angle reaches about $25^{\circ}$ at around $1.2$ Gyr 
and decreases quickly. 
For every case, the warp cannot be retained for longer than 4 Gyr.

\subsubsection{The Effect of the Incident Angle}

We define the incident angle $i$ as the angle between 
the orbital plane of the perturber and the equatorial plane of the disk galaxy. 
In previous sections, the incident angle $i$ was set to $90^{\circ}$ for all simulation runs. 
In reality, however, galaxies interact with no preference in angle,
and even for the same $M_{p}/M_{h}$ and $R_{ip}$, 
the strength and the lifetime of warps will vary according to different $i$ values.
We examine the effect of perturbers with various incident angles ranging from $i=0^{\circ}$ (A1) to $i=180^{\circ}$ (A9) 
on the disk galaxy during fly-by encounters (Table \ref{tbl-3}, the `A' runs). 

Early work by \citet{1972ApJ...178..623T} showed that 
prograde fly-by encounters induce a stronger host galaxy response than retrograde fly-by encounters. 
This phenomena is also well exhibited in our simulation results, 
showing well-developed spiral arms in one of the prograde models, A1 (bottom panels in Figure \ref{surface_density}). 
We refer to \citet{2008ApJ...683...94O} for detailed physical properties of spiral arms in interacting systems.

The fifth and bottom rows of Figure \ref{represent_fit} show the results of the A3 and A7 runs, 
for which $i=45^{\circ}$ and $135^{\circ}$ are used, respectively.  
At $t \sim 1.2 $ Gyr, warps reach their maximum 
and the overall pattern of evolution up to this point is almost identical to that of the simulation with $i=90^{\circ}$ (run A5 and IP2). 
A difference is found after reaching maximum warp. 
A warp angle with $i=90^{\circ}$ drops quickly in a few billion years 
and no sign of a warp remains after 4 Gyr. 
However, when $i=45^{\circ}$ or $135^{\circ}$, 
warps do not disappear until the end of our simulation. 

The rightmost column of Figure \ref{lifeinc} shows the averaged inclination and lifetime
of each `A' run in terms of the annulus.
The main lesson to be drawn from this result is that 
the incident angle plays a critical role in determining the angle and lifetime of warps, 
with the latter more closely related to the integration time of the encounter.
Fly-bys with more inclined incident angles have a longer integration time,
but their tidal forces are imposed mainly in the horizontal direction, 
not the vertical direction.
As a result, little or no disk warp is found for $i=0^{\circ}$ (A1) and $180^{\circ}$ (A9), 
i.e., with a perturber moving on the same plane as the host galaxy's disk. 
In contrast, fly-bys with $i=90^{\circ}$ (run A5) exert the largest vertical tidal force, 
but have the shortest integration time.
For this reason, the angle and lifetime reach their peaks 
at the midpoint between $i=0^{\circ}$ and $90^{\circ}$ (i.e., $i \simeq 45^{\circ}$) 
and between $i=90^{\circ}$ and $180^{\circ}$ (i.e., $i \simeq 135^{\circ}$), respectively.

\begin{figure*}[htbp]
\includegraphics[width=0.5\linewidth]{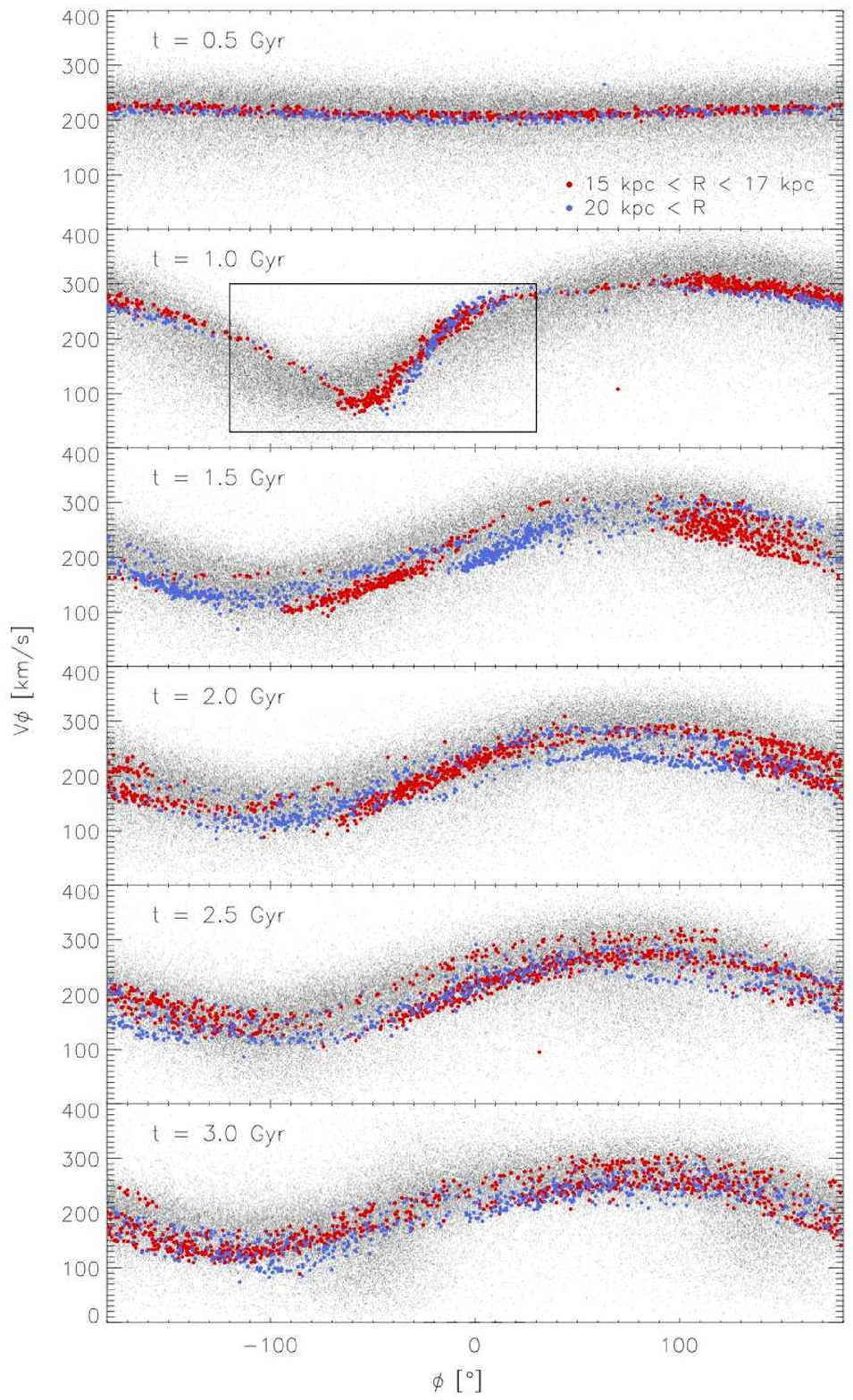}
\includegraphics[width=0.5\linewidth]{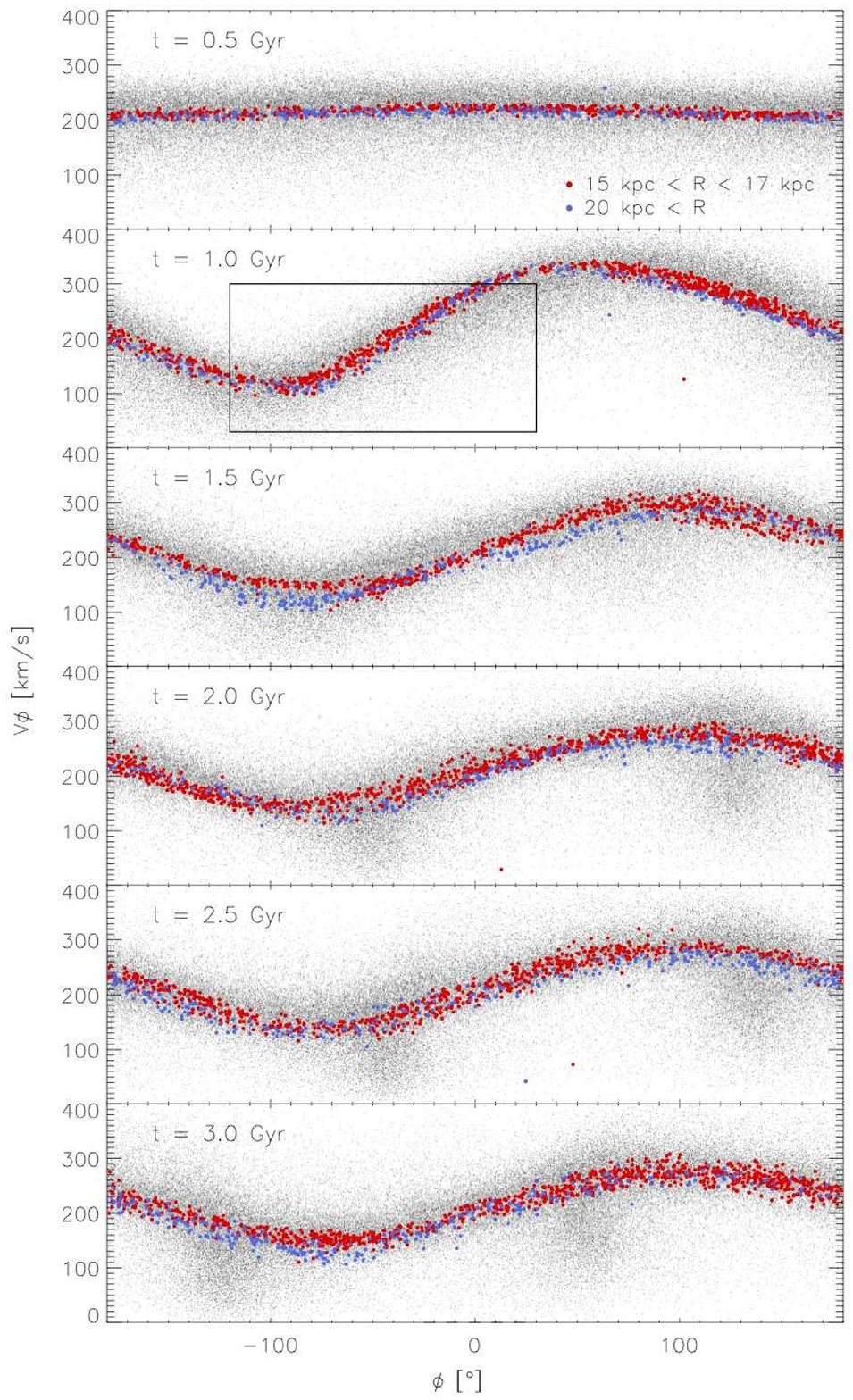}
\caption[]{Variations of the azimuthal velocities, $v_{\phi}$, as a function of the azimuthal angle ($\phi$) for all particles (gray), particles at 15 kpc $< R <$  17 kpc (red), and particles at $R > 20$ kpc (blue) in the A3 (left panels) and A7 (right panels) models. In each panel, $\phi = 0^{\circ}$  matches with the {\it x}-axis in our schematic view in Figure \ref{sview}.}
\label{velocity_variation}
\end{figure*}

\begin{figure*}[t]
\centerline{\includegraphics[width=0.95\linewidth]{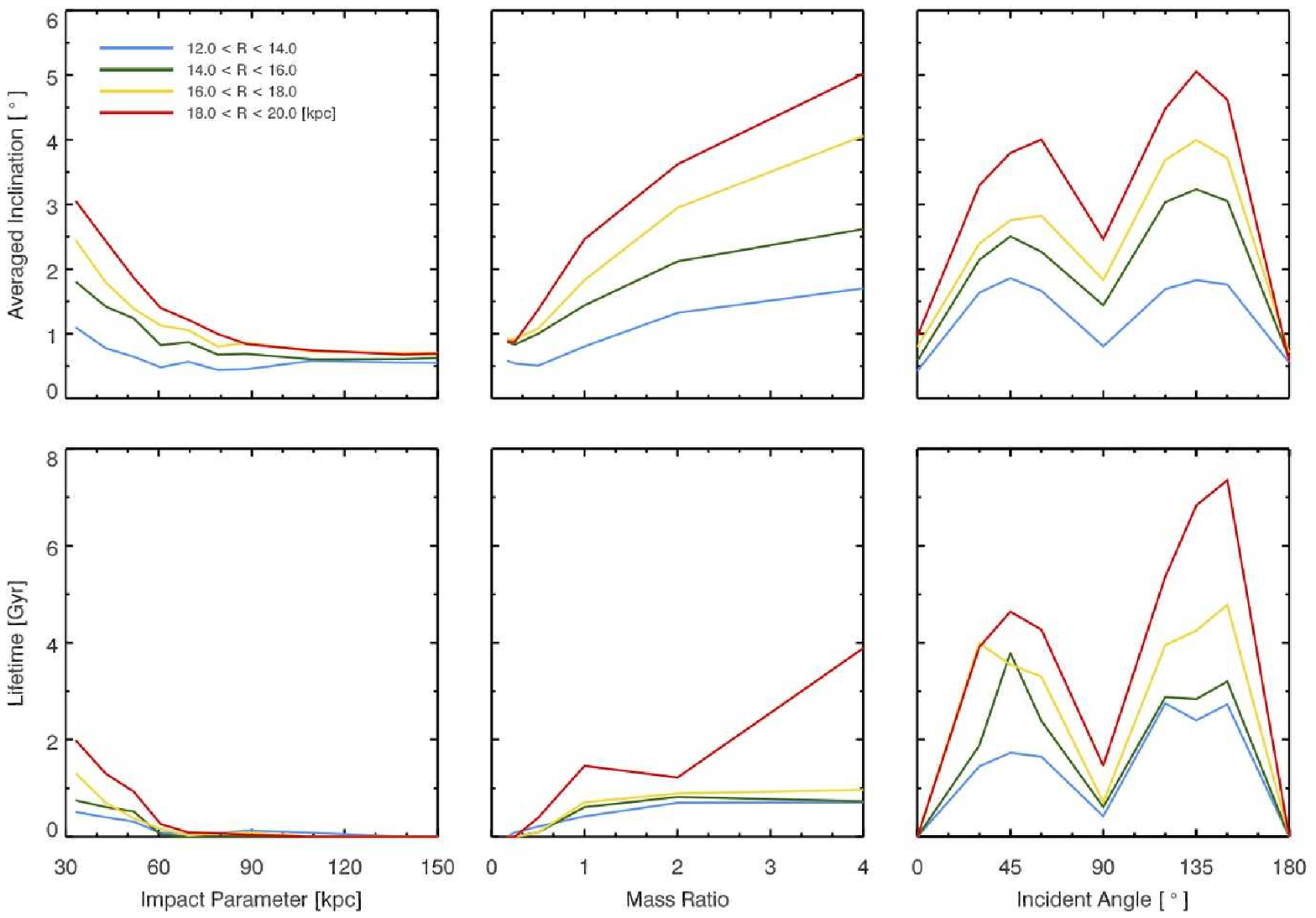}}
\caption[]{Warp amplitude (upper row) and lifetime (lower) for individual rings 
as functions of the impact parameter (left panels), mass ratio (middle), and incident angle (right).}
\label{relation}
\end{figure*}

It is important to note that there is a sizable disparity in warp angle and lifetime
between the $i=45^{\circ}$ and $135^{\circ}$ cases, 
in the sense that the $i=45^{\circ}$ case (A3) has a steeper slope of change in inclination than the $i=135^{\circ}$ case (A7). 
Incident angles could be treated the same way as
when the angle between the incoming path of the perturber and the equatorial plane is measured. 
However, the $i=45^{\circ}$ perturber follows the direction of the disk galaxy's rotation (prograde),
while the $i=135^{\circ}$ perturber moves in the opposite direction (retrograde).
For the prograde encounter, stellar particles of the host galaxy on the side closest to the perturber 
move in the same direction as the perturber, exerting tidal perturbations for a longer time.
We note that the lifetime of the A3 run is estimated to be shorter than that of the A7 run.
The outermost regions are disrupted severely,
but not in the form of galaxy warps,
due to the strong tidal force exerted on the prograde models.

We also find that the incident angle is closely related to the evolution of the line of node (LON). 
Figure \ref{pa} is the Briggs diagrams \citep{1990ApJ...352...15B}, 
showing the evolution of warps in terms of their amplitude and LON. 
In the diagrams, radial bins are spaced equally between 0 and 20 kpc (blue and red points, respectively) 
and the polar coordinate shows the warp angles ($3^{\circ}$, $6^{\circ}$, and $9^{\circ}$) of the LONs (concentric circles). 
Simulations show that the tip of a warp of first develops with respect to the direction of the perturber's incoming path.
As for the case of the A3 model run, the LON evolves in the direction of the disk's rotation for $\sim$ 0.5 Gyr
and turns its direction afterward,
whereas the LON of the A7 model evolves in the direction opposite to the disk rotation from the beginning.
The LON of the inner region advances faster than that of the outer regions. 
As a result, the diagrams gradually turn into leading spirals for both models.

Figure \ref{velocity_variation} depicts the variations of the azimuthal velocities of the A3 and A7 models
as a function of $\phi$.
Before an encounter, the azimuthal variation remains at $\sim$ 220 km s$^{-1}$. 
The orbits of particles are arranged in such a manner that 
the azimuthal velocities are symmetric with respect to the {\it x}-axis ($\phi$ = 0$^{\circ}$), 
the point where the perturber passes by most closely. 
The signature of the azimuthal velocity variation is apparent 
while the perturber approaches the host, reaching its maximum at {\it t} $\sim$ 1.0 Gyr.
The azimuthal velocities of the particles for the A3 and A7 models reach minimum values at $\phi \sim -60 ^{\circ}$ and $\phi \sim -110 ^{\circ}$, respectively.
(See the inside of small boxes in each panel of Figure \ref{velocity_variation} at {\it t} $\sim$1 Gyr).
These values of $\phi$ agree with the position angles of the warps we found in Figure \ref{pa},
implying that warps develop where the azimuthal velocity of galaxy is at a the minimum.
The difference between prograde and retrograde models is that the overall distributions are slightly shifted, 
which basically depends on the perturbers' path.
At {\it t} $\sim$1 Gyr, the velocity distribution in the outermost regions (red and blue dots) of the A3 models does not seem to follow the that of inner regions (gray),
showing a more shifted distribution due to the formation of strong spiral arms.

\subsubsection{Summary on the Effects of the Three Parameters}

A summary of all simulation runs 
in terms of the warp amplitude (upper row) and lifetime (lower row) as functions 
of three parameters---impact parameter (left panels), mass ratio (middle), and incident angle (right)
is provided in Figure \ref{relation}. 
The tendency of the warp's evolution can be summarized as follows: 
(1) the outer part of the disk is more affected by fly-by encounters,
(2) more massive perturbers and/or closer interactions trigger the formation of stronger, longer-lasting warps, and
(3) the perturber's incoming path matters, in that it determines 
how long the bending structure persists. 
For instance, 
in the case where $R_{ip}$ exceeds 100 kpc, a galactic interaction even with $M_{p}/M_{h}=1$ 
is unlikely to excite a warp. 
In the case of interactions with similar or more massive perturbers, 
relatively large warps are created, 
but $M_{p}/M_{h}\leq1/4$ is not able to excite warps. 
In the case of perturbers with $i=0^{\circ}$ and $180^{\circ}$, 
no visible warp is generated. 
Perturbers with $i\simeq45^{\circ}$ and $135^{\circ}$ excite the largest warps that persist for the longest time. 
Inclination angles and estimated lifetime decrease as $i$ approaches $90^{\circ}$, 
resulting in an `M'-shaped curve for the warp angle and a lifetime that is a function of the incident angle.

\section{Discussion}

\begin{figure*}[htbp]
\centerline{\includegraphics[width=\linewidth]{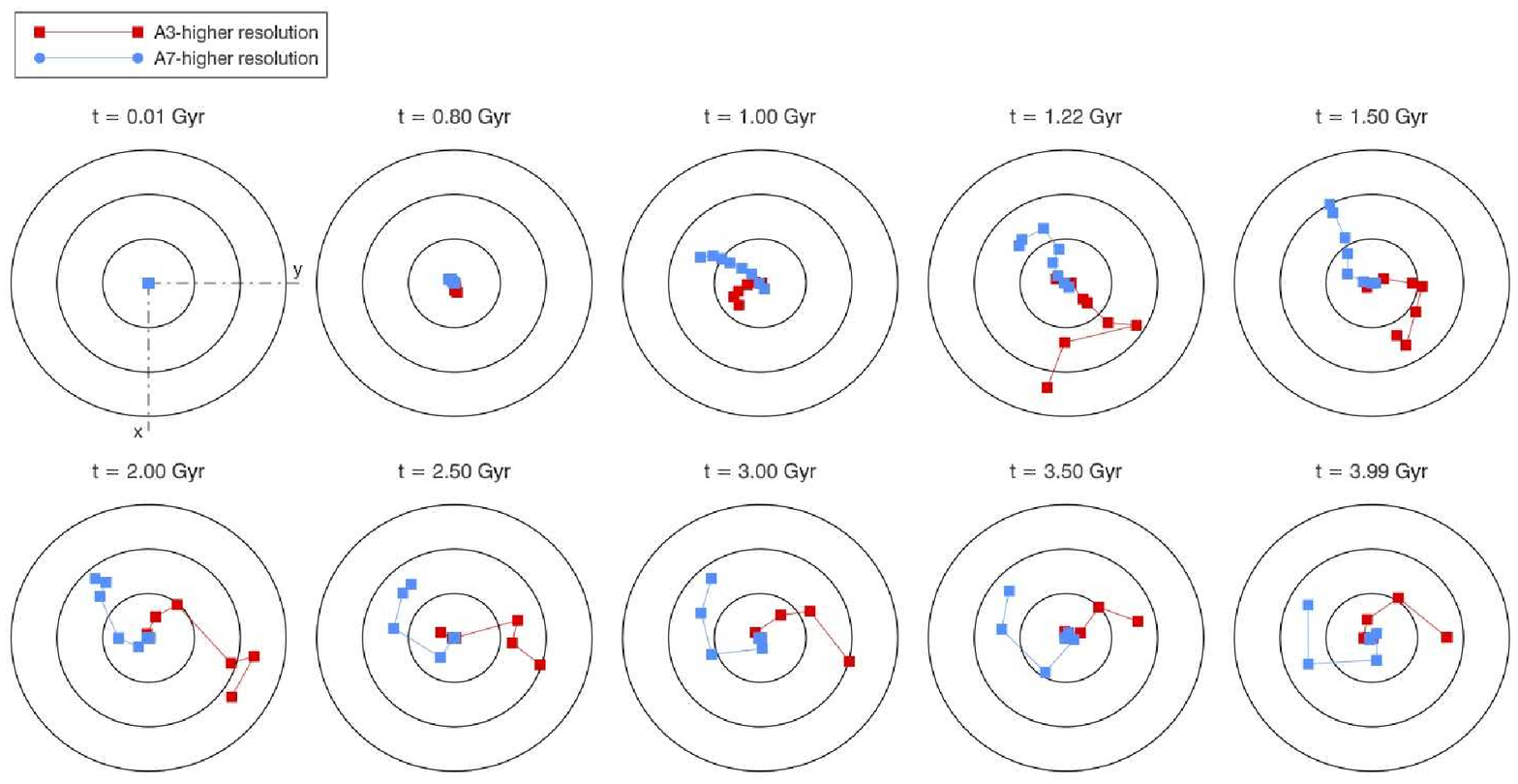}}
\caption[]{Same as Figure \ref{pa} but for the higher resolution model for the A3 (red) and A7 (blue) runs. The model contains one million disk particles, achieving ten times higher resolution than the intermediate models.}
\label{pa_high}
\end{figure*}

\subsection{The Effect of the Mass Resolution}
Using 100,000 particles in the disks might be inadequate to treat the subtleties of the disk edge phenomenon.
Disks at this resolution are subject to internal heating that increases the disk scaleheight and possibly induces artificial bending instabilities. 
Therefore, it is important to test for the numerical convergence of our results with higher resolution simulations.

In this section, we discuss the effect of particle resolution 
on the formation and evolution of galaxy warps induced by fly-by encounters, 
especially on the amplitude, position angle and its persistence. 
In previous runs, we used 100,000 particles for disks, 
ending up with a total number of around 1.4 million particles for the whole galaxy model. 

To test the sensitivity of galaxy warp evolution to numerical resolution, 
we conduct two additional experiments which are upgraded from the previous model simulation. 
Two models are exactly identical to the A3 and A7 runs aside from the fact that 
new higher-resolution sets now have one million particles for disks. 
Note that the number of other components of the model galaxy are also increased
(6.4 million, 55,000 and 185,000 particles for dark halo and the gas in the disks and the bulge, respectively). 
The number of particles used for the perturber in these new sets are 6.8 million.

Figure \ref{pa_high} shows Briggs diagrams for the higher resolution model A3 and A7  runs. 
A Briggs diagram is useful when comparing our intermediate and higher resolution models 
because we can easily visualize two major features,warp amplitude and position angle, simultaneously. 
Although there was a small deviation in the amplitude and position angle due to the random fluctuation, 
still we observe almost identical patterns in warp evolution between the intermediate (Figure \ref{pa}) and higher resolution simulation (Figure \ref{pa_high}). 
Thus, this suggests that particle resolution does not affect much the evolution trend of warps induced via fly-by encounters.

\begin{figure*}[htbp]

\includegraphics[width=0.5\textwidth]{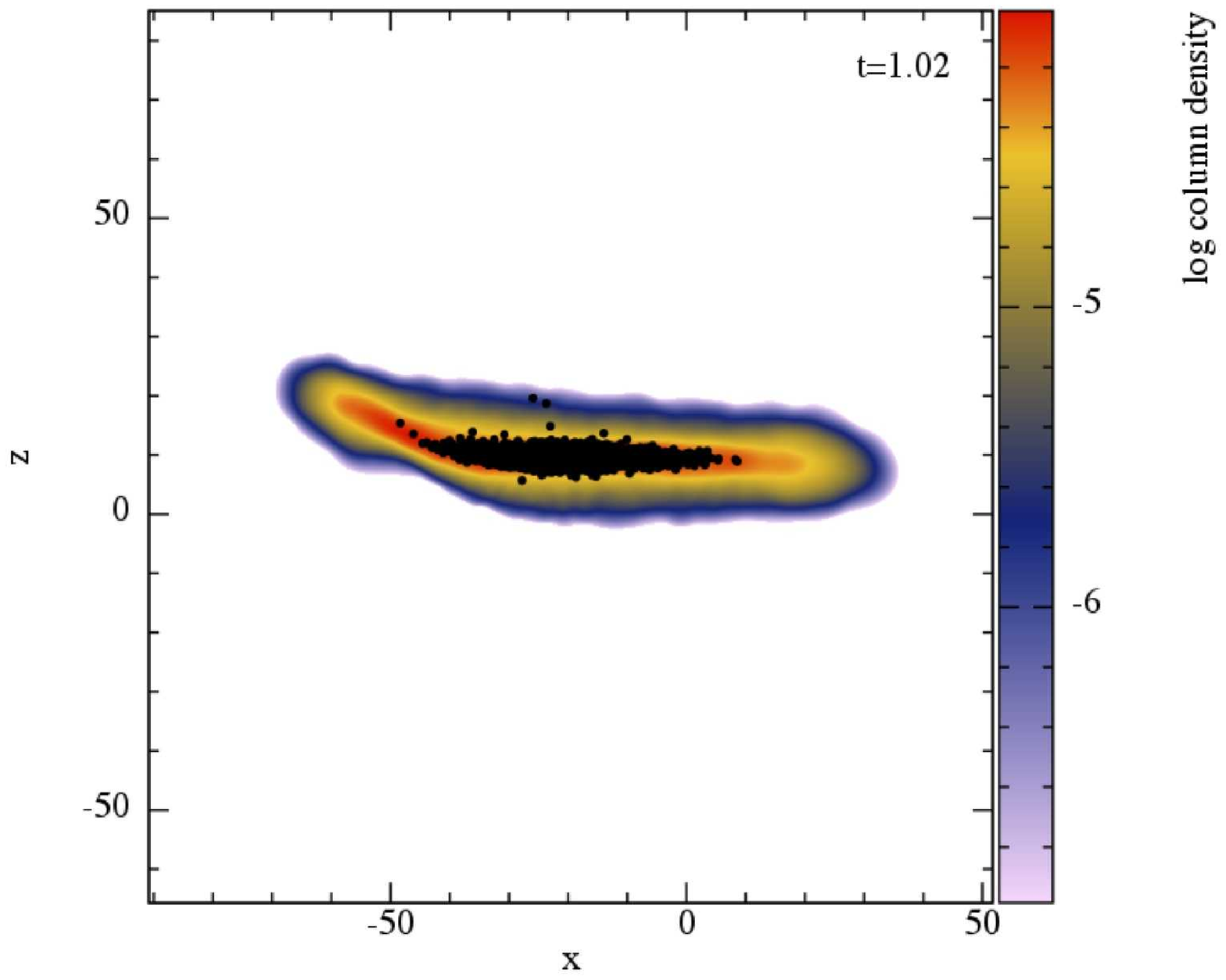}
\includegraphics[width=0.5\textwidth]{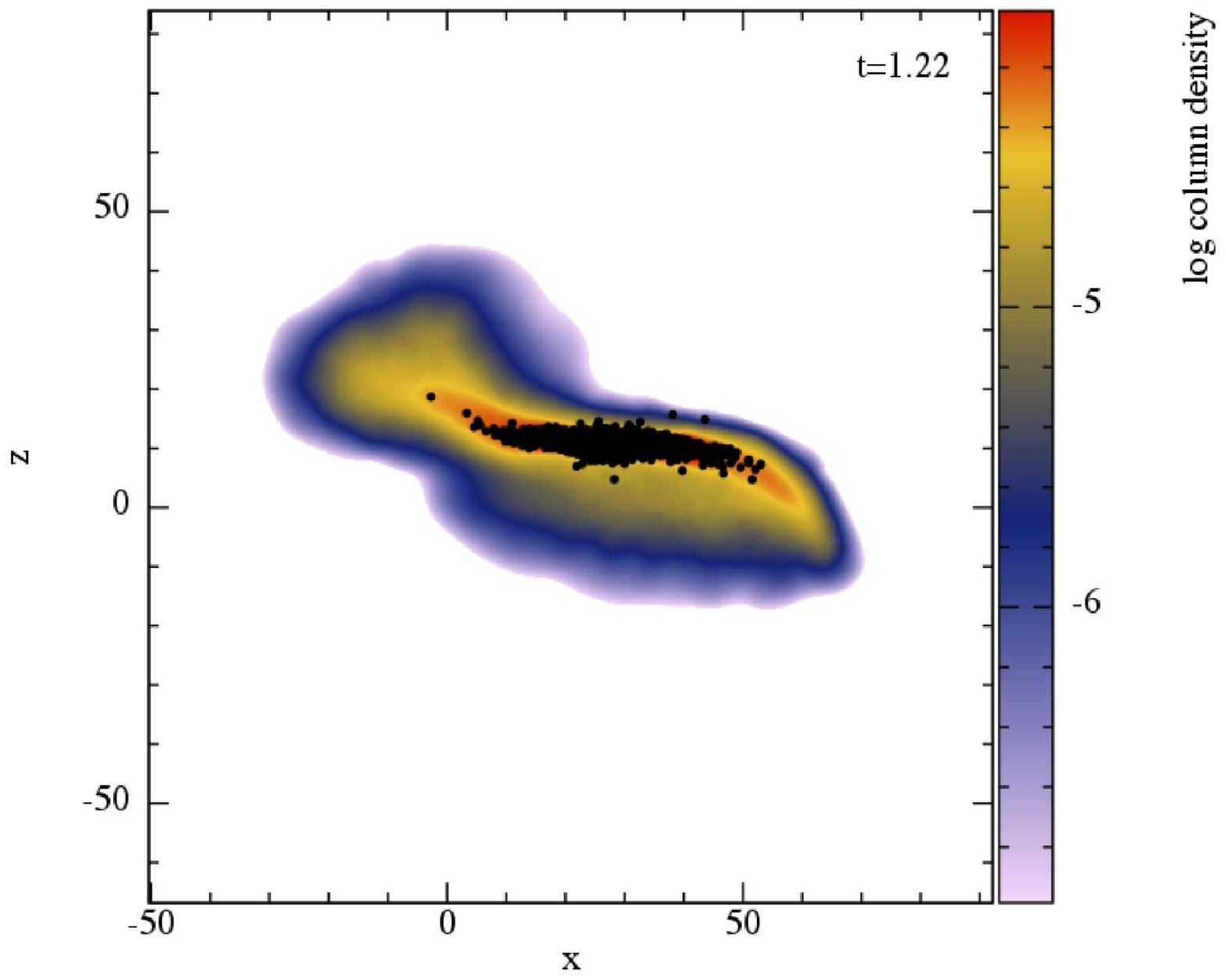}
\includegraphics[width=0.5\textwidth]{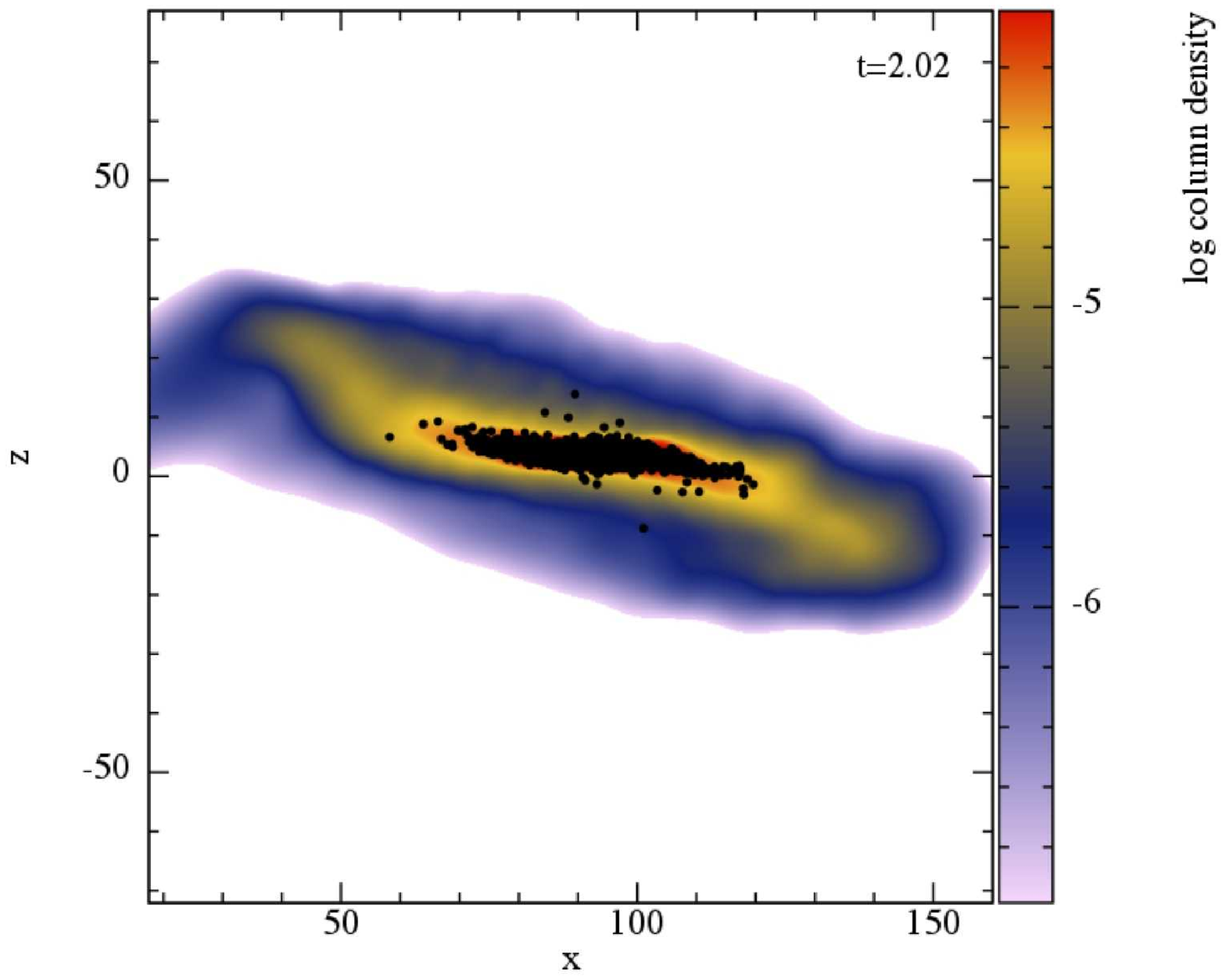}
\includegraphics[width=0.5\textwidth]{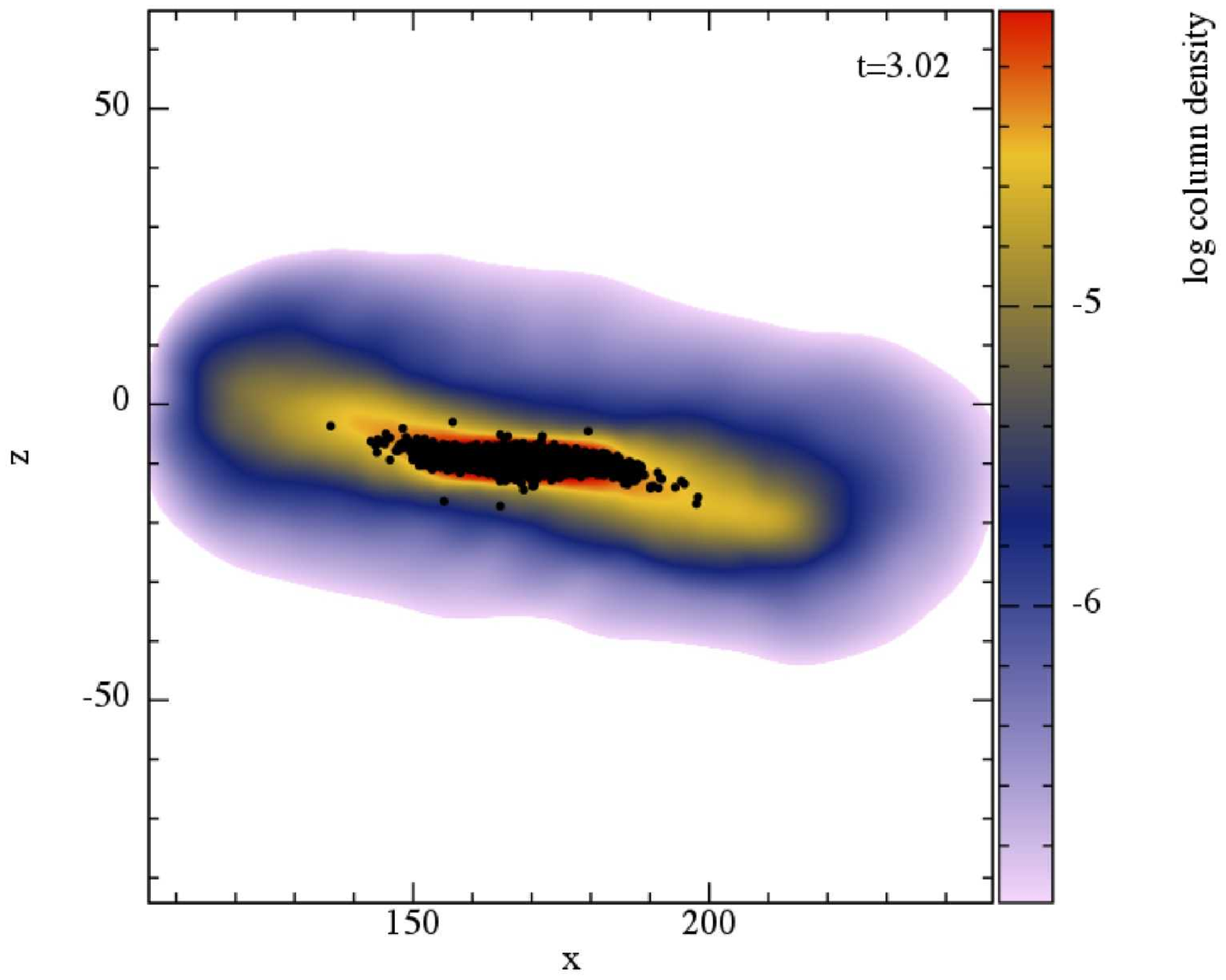}

\caption[]{Edge-on view of the projected gas particle number density with stellar particle distribution in black dots at $t\sim1.0\, $Gyr (top-left), $t\sim1.2\, $Gyr (top-right), $t\sim2.0\, $Gyr (bottom-left), and $t\sim3.0\, $Gyr (bottom-right). The color bar displays the level of the number density.}
\label{splash}
\end{figure*}

\subsection{Warps of Galaxies in the Field Environment}
We do not have a clear explanation for the origin of warps in field environments.
Both the Milky Way galaxy and M31, for example, are significantly warped, 
yet they show no sign of recent interaction with a large galaxy in the past few billion years.
Moreover, M33, the largest satellite of M31, is only 10 \% of the host, 
which may not be massive enough to create warps even during the interaction.
Interestingly, \citet{2006ApJ...641L..33W} showed that 
Magellanic Clouds can produce distortions in the dark matter halo of the Milky Way. 
They stated that the combined effect of these distortions and 
the tidal force exerted by Magellanic Clouds account for the creation of Galactic warp.

An interesting population of galaxies, the so-called backsplash galaxies, may provide a hint for warps in field galaxies.
Backsplash galaxies are individual galaxies that once visited the core regions of a galaxy cluster, 
deep within the cluster potential $\sim$ 0.5 $R_{vir}$ and rebounded up to $\sim$ 2.5 $R_{vir}$
so that they are now located in the outskirts of the cluster, 
obscuring the definition of cluster and field environments \citep{2004A&A...414..445M, 2005MNRAS.356.1327G}.
Several observational studies have confirmed the existence of this new population in galaxy clusters \citep{2004A&A...418..393S,2006MNRAS.366..645P},
and backsplash galaxies even in isolated galaxy clusters that are free from recent cluster$-$cluster merger activity  \citep{2011MNRAS.411.2637P}.
The backsplash galaxies in the outskirt of clusters should have high chance of fly-by interactions 
and thus of exhibiting warp phenomena.

\subsection{The Effect of Triaxial Halos}
It has been suggested that warps in disks might survive for a long time
if halos are at least three times more massive than the disk and
nearly spherical (axial ratio $<$ 1.2),
obviously triggering no precession of disks \citep{1979ApJ...230..736T,1983IAUS..100..177T}.
Thus, this basic idea is consistent with our results.

We prefer spherical {\it live} halos to a static fixed potential 
so as not to underestimate the effect of dynamical friction 
on disks residing on halos when halos overlap during the encounter.
However, as many numerical studies demonstrate, 
the shape of the dark matter halos are, in fact, triaxial 
\citep{1991ApJ...378..496D,2002ApJ...574..538J,2005ApJ...627..647B,2006MNRAS.367.1781A,2012ApJ...748...54Z},
favoring flattened dark halos.

The effect of a triaxial halo on the evolution of warps in galaxies with disk particles in a fixed halo potential has been examined. 
One of the key results is that warps in triaxial halos show oscillatory behavior. 
The triaxiality of halos also cause the differential precession and nutation between the inner and outer regions of the disks, 
attenuating or fluctuating the magnitude of the warps \citep{2000MNRAS.311..733I,2009ApJ...696.1899J}.
It is a well-known fact that the main difficulty in maintaining coherent warps of galaxies 
is the problem of differential precession of inclined orbits in the combined flattened potential of the disk and asymmetric dark halo.

Unfortunately, our scheme is currently incapable of generating a stable triaxial halo and galaxy with live particles.
It is beyond the scope of this paper to run fly-by simulations 
in order to demonstrate the effect of the flattened halo on the evolution of warps.
Given these limitations, our aim in the present study is to understand the fundamental effects of galactic fly-by encounters 
on the formation and evolution of disk warps.

\subsection{Stellar Disk versus Extended Gas Disk}

The galaxy models in this study do not contain particles that describe extended \HI\ gas disks. 
However, extended \HI\ warps are universal, 
and their amplitudes are usually greater than those of stellar warps \citep[e.g.,][]{1998A&A...337....9R}. 
\citet{1996AJ....111.1505C} showed that warps in both optical and neutral hydrogen gas 
share similar characteristics, supporting the idea that gravitational force 
plays a crucial role in the formation of warped galaxies \citep{1992ARA&A..30...51B}.
To ensure that the fly-by mechanism also works for the formation of gaseous warps, 
we perform a single run that includes the extended \HI\ gas containing 10 \% of the total gas mass.

Four snapshots of the projected particle number density 
($t$\,$\simeq$\,1.0, 1.2, 2.0 and 3.0 Gyr) 
of the simulated gas distribution along with the stellar distributions
are demonstrated in Figure \ref{splash}. 
The simulation parameters are chosen to be identical to the A2 model.  
In line with observations, the gas distribution of our model follows that of stellar disks
and an extended \HI\ warp develops with a higher amplitude than that of stellar warp. 
Details of gas behavior will be presented in a separate paper in this series.

\begin{figure*}[]
\centerline{\includegraphics[width=\linewidth]{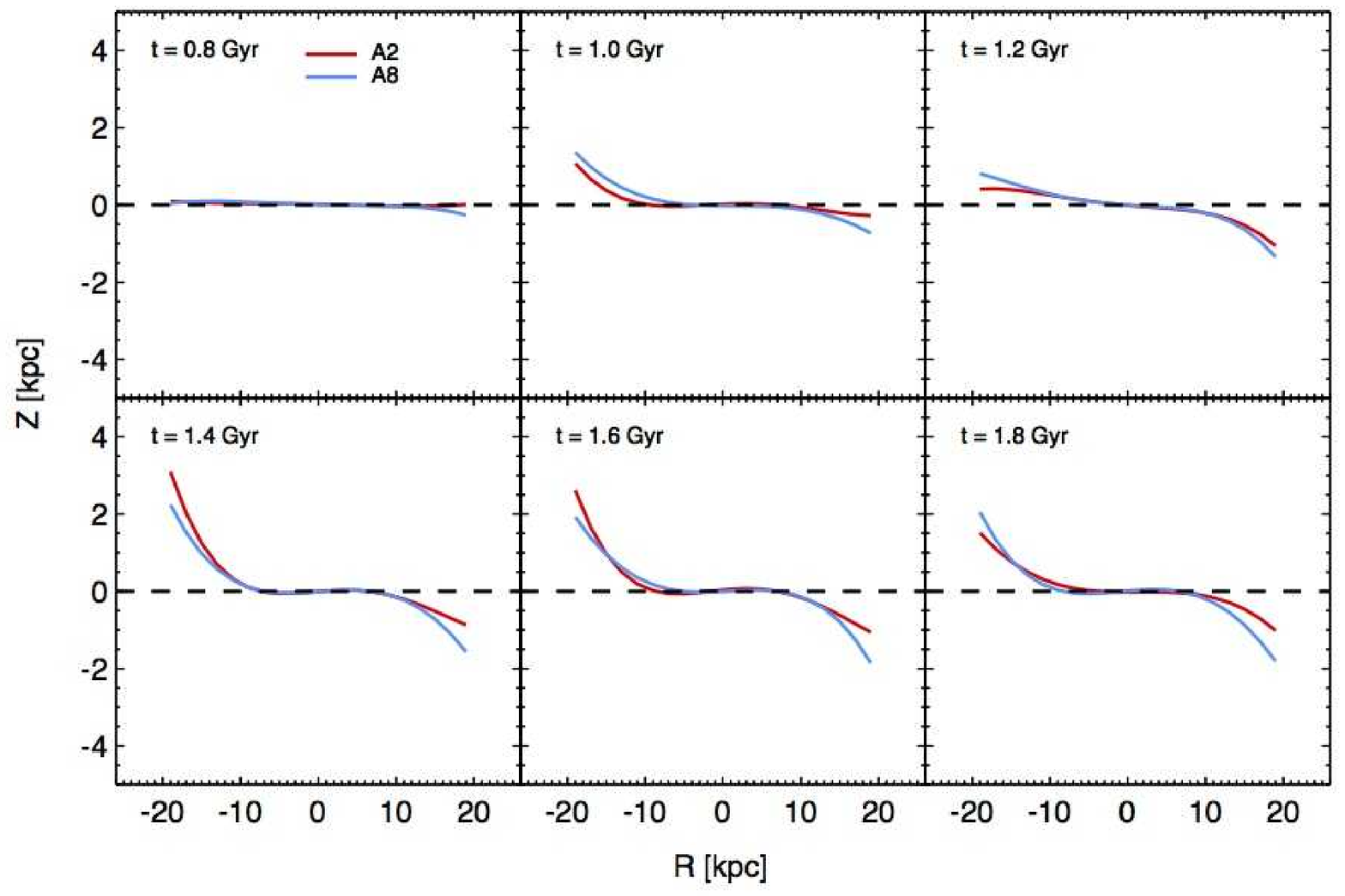}}
\caption[]{Fiducial lines of particles of galaxies on the edge-on view for models A2 (red) and A8 (blue) at different epochs. Before fiducial lines are drawn, we rotate each galaxy to an angle where the warp amplitudes are highest. Therefore, the angles are measurable at each side. Black dotted lines depict reference points.}
\label{asym2}
\end{figure*}

\begin{figure*}[htbp]
\includegraphics[width=0.5\textwidth]{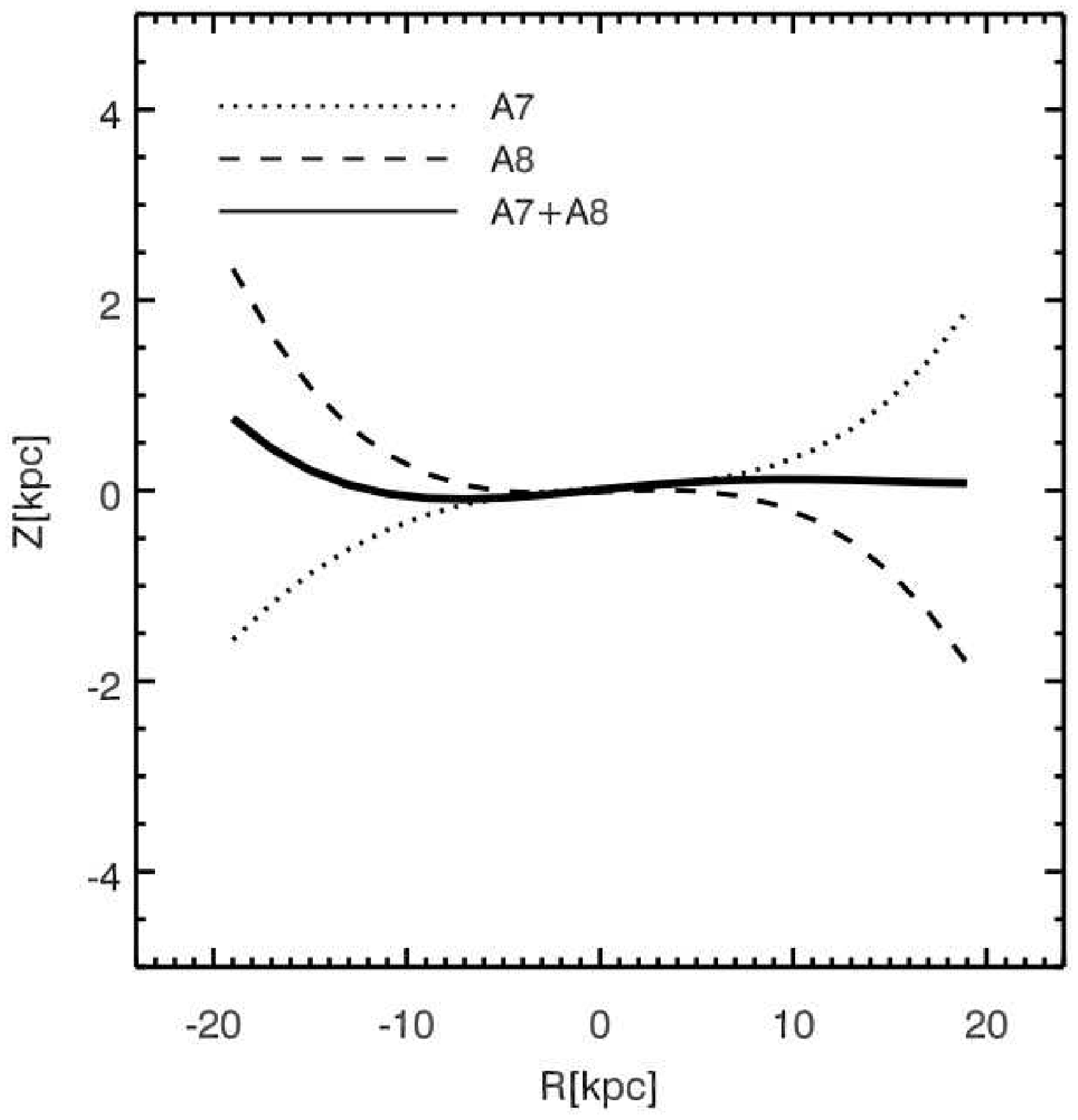}
\includegraphics[width=0.5\textwidth]{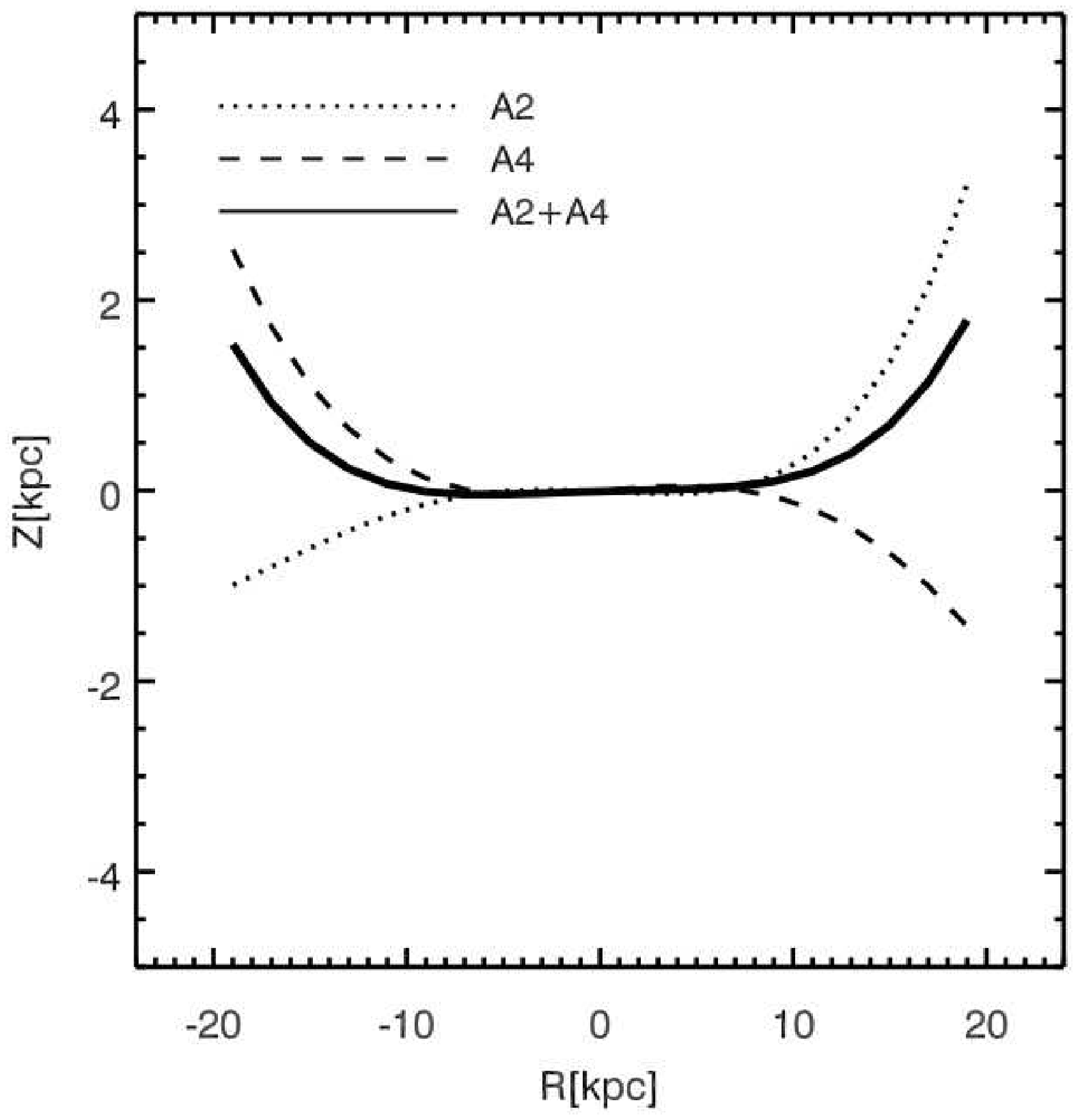}
\caption[]{Examples of modeled L- and U-shaped warps created by superposition of non-symmetric S-shaped warps. The dotted lines and dashed lines are drawn in the same manner that Figure \ref{asym2} is depicted. The solid line represents the combination of two models. }
\label{asym_sum}
\end{figure*}

\subsection{Asymmetric Warps and the Possible Origin of U- and L-shaped Warps}

A considerable number of warped galaxies show noticeable asymmetries 
\citep{1994A&A...290L...9R,1998A&A...337....9R,2002A&A...391..519C,2006NewA...11..293A}. 
The degree of asymmetry can be quantified 
by the difference in warp magnitude at each side, using 

\begin{equation}
A_{\alpha}=|\alpha_1-\alpha_2|\,\,\,,
\end{equation}
where $\alpha_1$ and $\alpha_2$ are the warp angles of one side and the opposite side, respectively
\citep{2006NewA...11..293A}.
Most models show an asymmetry of $A_{\alpha}<1^{\circ}$ for the majority of the time 
and $1^{\circ}<A_{\alpha}<2^{\circ}$ for a short period of time right after the interaction.
Among our galaxy models, the A2, A3 and A4 runs show 
the most prominent asymmetric warps ($A_{\alpha}>6^{\circ}$ at $t=1.4$ Gyr).

An example of a non-symmetric warp in our runs (A2 in red) 
along with a relatively symmetric warp (A8 in blue) for comparison
is shown in Figure \ref{asym2}. 
As a perturber is still approaching ($t < 1$\,Gyr), 
one side of the disk (the side nearest the perturber) bends first. 
After the perturber passes by ($t >1$\,Gyr), the other side of the disk starts to bend, 
resulting in an S-shaped warp.
For a few billion years, some warps in our models develop strong asymmetry. 
Later on, the stronger side of the warp descends to the same level as the other side of the warp, 
resulting in a normal symmetric warp.
We note that once a non-symmetric stellar warp is present, 
the extended gas disk has a larger $A_{\alpha}$ 
and the asymmetry lasts longer than the stellar counterpart.

In Figure \ref{asym_sum}, we speculate that the observed U- and L-shaped warps are 
geometric superpositions of two (or more) non-symmetric S-type warps. 
More than one successive fly-by, each with a different incident angle,
can cause two S-types to be superimposed.
Multiple fly-bys may include interactions with a satellite on a highly elongated orbit
Suppose that an intruder galaxy perturbs a disk, producing a symmetric warp, 
and that the warped galaxy experiences another fly-by encounter generating an asymmetric warp.
In this case, the initial S-type can be modified by the latter incident,
leading to an L-shaped (left panel) or U-shaped warp (right panel).
Observations show that the minority ($\sim$\,30\,\%) of stellar warps are U- and L-types 
while the S-type accounts for $\sim$\,70\,\% of stellar warps \citep{2006NewA...11..293A}.
Because multiple interactions with proper incident angles are required 
to form U- and L-shaped warps under the superposition scheme, 
the fraction of these types should naturally be less than that of the dominant S-types.



\section{Conclusions}

We performed a set of {\it N}-body simulations with live halos 
to investigate the morphological and kinematical evolution of disk galaxies 
that experience fly-by encounters. 
We found that warps can be excited by impulsive encounters 
and can be sustained for a few billion years. 
The magnitudes of the warps reach maximum values 
from a few degrees up to $\sim 25^{\circ}$, 
and warps survive for a few billion years
depending on three major parameters:
(1) minimum distance, (2) mass ratio, and (3) incident angle.
Our results coincide with the fact that most optical warps are weak,
and confined to the outer parts of the galaxy.
While the maximum amplitude is tied up with all parameters listed above, 
the warp lifetime is determined mainly by the incident angle of the perturber
because it affects the integration time and the direction of the force exerted.
In addition, the tip angle of a warp first develops with respect to the direction of the incident angle 
where the azimuthal velocity of the galaxy is at a minimum,
and then evolves in the direction opposite to the disk rotation after all.

Some models show strong non-symmetric warps ($A_{\alpha}>6^{\circ}$) at a certain period of time. 
This suggests that the superposition of asymmetric warps, 
created by two successive fly-by encounters 
or even multiple interactions with a satellite on a highly elongated orbit, 
generates U- and L-shaped warps. 
If this is the case, the amplitudes of most U- and L-shaped warps should be smaller than that of S-shaped warps, 
which is consistent with observations.

We also briefly described how \HI\ disks forms a bending structure 
similar to that of stellar disks in response to fly-by encounters,
except that the magnitude of the \HI\ warps is greater than that of stellar warps. 
How the gas components react to fly-bys is also an important issue to consider. 
Details of gas behavior will be presented in a subsequent paper in this series.


\acknowledgments
S.-J.Y. acknowledges support from Mid-career Research Program (No. 2012R1A2A2A01043870) 
through the National Research Foundation (NRF) of Korea, 
and support by the NRF of Korea to the Center for Galaxy Evolution Research (No. 2013-8-1583) 
and by the Korea Astronomy and Space Science Institute Research Fund 2013--2014. 
S.S.K's work was supported by the Mid-career Research Program (No. 2011-0016898)
through the National Research Foundation (NRF) grant funded by the Ministry of
Education, Science and Technology (MEST) of Korea.



\end{document}